\documentclass[a4paper,11pt,oneside]{article}
\usepackage[margin=2.5cm]{geometry}
\usepackage{amsthm, amsmath, amssymb}
\usepackage{cite}

\usepackage[toc,page]{appendix}
\usepackage{titling}

\usepackage{booktabs}
\usepackage{enumitem}

\usepackage[utf8]{inputenc}

\usepackage{subcaption}

\usepackage{tikz}
\usetikzlibrary{arrows,shapes,trees, arrows.meta}

\newtheorem{theorem}{Theorem}
\newtheorem{corollary}[theorem]{Corollary}


\title{\vspace{1pt} Stabilization Time in Minority Processes \vspace{90pt}}

\author{\Large Pál András Papp \footnotesize \vspace{8pt} \\ ETH Zürich \vspace{2pt} \\ apapp@ethz.ch
\and \and \Large Roger Wattenhofer \footnotesize \vspace{8pt} \\ ETH Zürich \vspace{2pt} \\ wattenhofer@ethz.ch}

\date{}

\begin{document}

\begin{titlingpage}
  \maketitle
	\vspace{130pt}
  \begin{abstract}
		{We analyze the stabilization time of minority processes in graphs. A minority process is a dynamically changing coloring, where each node repeatedly changes its color to the color which is least frequent in its neighborhood. First, we present a simple $\Omega(n^2)$ stabilization time lower bound in the sequential adversarial model. Our main contribution is a graph construction which proves a ${\Omega}(n^{2-\epsilon})$ stabilization time lower bound for any $\epsilon>0$. This lower bound holds even if the order of nodes is chosen benevolently, not only in the sequential model, but also in any reasonable concurrent model of the process.}
  \end{abstract}

	%This construction works in a large class of models: adversarial, random, benevolent, sequential as well as concurrent.
	%The bound obtained from this construction holds not only in the sequential model, but also in any reasonable concurrent model of the process, even if the order of nodes is chosen benevolently.
	
	\vspace{5pt}
	
\end{titlingpage}

\setcounter{page}{2}

\pagenumbering{arabic}

\section{Introduction}

If you google ``bad wifi'', one advice you will get for sure is to choose the least crowded frequency in order to minimize interference with your neighbors. Unfortunately, this least crowded frequency may change again if some of your neighbors do the same.

Frequency allocation is a familiar example of \textit{minority processes} in graphs: given a graph, a set of colors, and an initial coloring of the nodes with these colors, a minority process is a process where each node, when given the chance to act, modifies its color to a color that has the smallest number of occurrences in its neighborhood. This results in a dynamically changing coloring, which is essentially a form of distributed automata. Minority processes arise in various fields of economics \cite{KPRanticoor} or social science \cite{applic4} when players are motivated to differentiate from each other, but they also emerge in cellular biology \cite{applic3} or crystallization mechanics \cite{applic1, applic2}.

%These processes are essentially a form of distributed automata, which arise in various fields of economics \cite{KPRanticoor} or social science \cite{applic4} when players are motivated to differentiate from each other, but also emerges in cellular biology \cite{applic3} or crystallization mechanics \cite{applic1, applic2}.

A minority process is said to stabilize when no node has an incentive to change its color anymore. The aim of the paper is to understand how long it takes until such a minority process reaches a stable state.
We study the process in several different models, some of them sequential, some concurrent.
In sequential models, when only one node at a time can change its color, stabilization time depends on the choice of the order of nodes. Hence, the model can further be subdivided into three cases, depending on whether the order of acting nodes is specified benevolently (trying to minimize stabilization time), adversarially (trying to maximize stabilization time), or randomly.

On the other hand, in concurrent models, multiple nodes are allowed to switch their color at the same time. However, if two (or more) neighboring nodes continuously keep on forcing each other to switch their color, the system may never stabilize. The simplest such example is a graph of two connected nodes that have the same initial color, and keep on switching to the same new color in every step. We also study concurrent models that exclude this behavior, as it is unrealistic in many application areas where neighbors are unlikely to switch at the exact same time.

In any model where simultaneous neighboring switches are excluded, it is easy to prove a $O(n^2)$ upper bound on stabilization time for minority processes. Initially, some (maybe even all) of the at most $O(n^2)$ edges in the graph are monochromatic (i.e., they have a \textit{conflict}). When a node switches its color to the minority color in its neighborhood (but its neighbors do not change color in the same step), then the number of conflicts on the adjacent edges strictly decrease. Since the original number of conflicts is $O(n^2)$ and the overall number of conflicts decreases by 1 at least in each step, the number of steps is limited to $O(n^2)$.

However, this raises a natural question: are there example graphs that exhibit this naive upper bound? Or is there a significantly lower (e.g. linear) upper bound on stabilization time in some models? While these questions are already answered for the ``dual'' problem of majority processes (when nodes switch to the most frequent color in their neighborhood), for the case of minority processes, they have remained open so far.

%While for the ``dual'' problem of majority processes, when nodes switch to the most frequent color in their neighborhood, these questions are already answered, for the case of minority processes, they have remained open so far.

% when nodes aim to coordinate with each other
% when nodes want to not differentiate from but coordinate with their neighbors
% when nodes switch to the most frequent color in their neighborhood

The main contributions of the paper are constructions that prove lower bounds on stabilization time of minority processes. As a warm-up, we present a simple example in Section \ref{sec:seqadv} which shows that in the sequential adversarial model, stabilization may take $\Theta(n^2)$ steps. Our main result is a construction proving that stabilization can also take superlinear time in the sequential benevolent case. We first present a graph and an initial coloring in Section \ref{sec:seqben} where any selectable sequence lasts for $\Omega(n^{3/2})$ steps. Then in Section \ref{sec:recurse}, we outline how a recursive application of this technique leads to a stabilization time of $\Omega(n^{2-\epsilon})$ for any $\epsilon>0$, almost matching the upper bound of $O(n^2)$. This is an interesting contrast to majority processes, where stabilization time is bounded by $O(n)$ in the benevolent case. Furthermore, our construction shows that this almost-quadratic lower bound holds not only in the sequential model, but also in any reasonable concurrent setting.

\section{Related work}

While there is a wide variety of results on both minority and majority processes, majority processes have been studied much more extensively. Recently, \cite{majority} has shown that stabilization time in majority processes can be superlinear both in the synchronous model, and in the sequential model if the order is chosen by an adversary. However, \cite{majority} has also shown that stabilization always happens in $O(n)$ time in the sequential benevolent model. In case of majority processes in weighted graphs, a $2^{\Theta(n)}$ lower bound on stabilization time was also shown in \cite{majorityW}.

Other aspects of majority processes have also been studied thoroughly, especially in the synchronous model. Results on majority processes include basic properties \cite{Goles, Winkler}, their behavior on random graphs \cite {MajOther2, Ahad2018}, complexity results on determining stabilization time \cite{votingtime}, minimal sets of nodes that dominate the process \cite{Fazli, MajOther1}, the existence of stable states in the process \cite{approx0, SGPclass1, SGPclass2, SGPclass3, SGPclass4, SGPclass5, SGPsurvey}, and modified process variants \cite {certainityMaj, switchOnce}.

In contrast to this, the dynamics of minority processes has received less attention. The stabilization of minority processes has only been studied in special classes of graphs, including tori, cycles, trees and cliques \cite{CA1, CA2, CA3}. These studies are mostly conducted only in the synchronous or the sequential random model. More importantly, these results study a different variant of the minority process, which considers the closed neighborhood of nodes, and thus can result in significantly larger (possibly exponential) stabilization time, even in the unweighted case. An experimental study of the processes on grids is also available in \cite{CA1}.

In weighted graphs, it has recently been shown in \cite{stacs} that stabilization of minority processes can take $2^{\Theta(n)}$ steps in various models, matching a straightforward exponential upper bound in the weighted case. However, the constructions of \cite{stacs} use exponentially large node or edge weights to obtain these results; as such, the same techniques are not applicable in the unweighted case.

Besides these studies on the dynamics of the process, there are also numerous theoretical results on stable states in minority processes. These include complexity results on deciding the existence of different stable state variants \cite{KPRanticoor, approx0}, characterization of infinite graphs with a stable state \cite{noUGP, aharoniUGP, UGPrayless}, and analysis of price of anarchy in such states as local minima \cite{KPRanticoor, PoAframework}. In the work of \cite{hedetniemi}, it is also shown that slightly modified minority processes, based on distance-2 neighborhood of nodes, can provide better local minima at the cost of larger (but still polynomial) stabilization time.

However, in contrast to majority processes, the stabilization time of minority processes in general unweighted graphs has remained unresolved so far.

\section{Definitions and background}

\subsection{Models}

In the paper, we primarily focus on the following models:
\vspace{4pt}
\renewcommand{\theenumi}{\Alph{enumi}}
\begin{enumerate}
	\item \textbf{Sequential adversarial:} In every step, only one node switches. The order of nodes is specified by an adversary who maximizes stabilization time.
	\item \textbf{Sequential benevolent:} In every step, only one node switches. The order is specified by a benevolent player who minimizes stabilization time.
	\item \textbf{Independent benevolent:} In every step, the benevolent player is allowed to choose any independent set of switchable nodes, and switch them simultaneously.
	\item \textbf{Free benevolent:} In each step, the benevolent player is allowed to choose any set of switchable nodes, and switch them simultaneously.
\end{enumerate}
\vspace{4pt}
\noindent
However, our lower bounds extend to a range of other popular models:
\vspace{4pt}
\begin{enumerate}
\setcounter{enumi}{4}
	\item \textbf{Concurrent synchronous:} In every step, all switchable nodes switch simultaneously.
	\item \textbf{Sequential random:} In every step, only one node switches, chosen uniformly at random among the switchable nodes.
	\item \textbf{Concurrent random:} In every step, every switchable node switches with probability $p$, independently from other nodes.
\end{enumerate}

\begin{figure}
\centering
\captionsetup{justification=centering}
	\scalebox{.9}{\begin{tikzpicture}

\draw[gray, thick, arrows=stealth-stealth] (0pt,0pt) -- (0pt,100pt);
\draw[gray, thick, arrows=stealth-stealth] (-60pt,20pt) -- (120pt,20pt);

\node[anchor=north] at (0pt,2pt) {\footnotesize \textit{sequential}};
\node[anchor=south] at (0pt,100pt) {\footnotesize \textit{concurrent}};
\node[anchor=north] at (-60pt,19pt) {\footnotesize \textit{adversarial}};
%\node[anchor=north] at (-60pt,9pt) {\footnotesize \textit{moveset}};
\node[anchor=north] at (120pt,19pt) {\footnotesize \textit{benevolent}};
%\node[anchor=north] at (120pt,9pt) {\footnotesize \textit{moveset}};

\draw[fill=black] (0pt,20pt) circle (0.5ex);
\draw[fill=black] (-45pt,20pt) circle (0.5ex);
\draw[fill=black] (45pt,20pt) circle (0.5ex);
\draw[fill=black] (0pt,50pt) circle (0.5ex);
\draw[fill=black] (0pt,80pt) circle (0.5ex);
\draw[fill=black] (70pt,50pt) circle (0.5ex);
\draw[fill=black] (95pt,80pt) circle (0.5ex);

\node[anchor=south] at (-39pt,20pt) {\normalsize A};
\node[anchor=south] at (51pt,20pt) {\normalsize B};
\node[anchor=south] at (75pt,50pt) {\normalsize C};
\node[anchor=south] at (101pt,80pt) {\normalsize D};
\node[anchor=south] at (8pt,80pt) {\normalsize E};
\node[anchor=south] at (8pt,50pt) {\normalsize G};
\node[anchor=south] at (8pt,20pt) {\normalsize F};
	
\end{tikzpicture}}
	\vspace{-5pt}
	\caption{Properties of the listed models}
	\label{fig:models}
	\vspace{-3pt}
\end{figure}
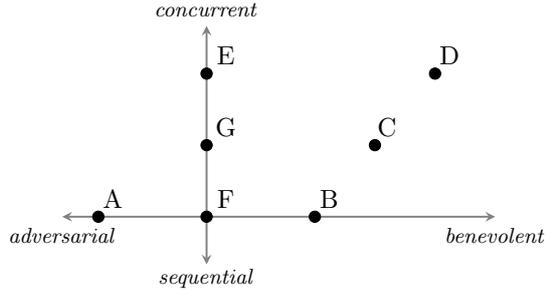

An intuitive illustration of these models is shown in Figure \ref{fig:models}. The vertical axis shows how concurrent a model is, the horizontal shows how wide is the set of opportunities it grants the player to speed up / slow down stabilization. In the case of majority processes, models A and E are shown to take superlinear time to stabilize for some graphs, but model B always stabilizes in linear time \cite{majority}. However, we prove that for minority processes, even model B can take superlinear time. Models C and D grant even wider sets of possible (concurrent) moves for the benevolent player, which may drastically reduce the number of steps in some cases; however, we show that the same lower bound holds even if such moves are available.

Note that models A, B, C and F exhibit a natural $O(n^2)$ upper bound on stabilization time, as the overall number of conflicts decreases in each step by at least 1. On the other hand, models D, E or G may allow neighboring nodes to switch at the same time, and thus in these models, some nodes may keep on endlessly changing colors. However, our constructions specifically ensure that connected nodes are never switchable at the same time, and thus for these particular graphs, the process stabilizes in any of the models.

Through most of the analysis in the paper, we focus on the sequential models. We first show a simple construction with $\Theta(n^2)$ stabilization time in model A. We then present a more complex construction to first show $\Omega(n^{3/2})$, and then $\Omega(n^{2-\epsilon})$ stabilization time in model B. It then follows from a few observations that these latter constructions also have the same stabilization time in models C and D. Since model D provides the widest set of opportunities from all models, this implies the same lower bound for each of the listed models.

\subsection{Preliminaries}

\vspace{-0.5pt}

Throughout the paper, we consider simple, unweighted, undirected graphs. Graphs are denoted by $G$, their number of nodes by $n$, and the maximum degree in the graph by $\Delta$.

Given a graph $G$ on the vertex set $V$, an \textit{independent set} is a subset of $V$ such that no two nodes in this subset are connected. A \textit{coloring} of the graph with $k$ colors is the assignment of one of the colors (numbers) from $\{1, 2, ..., k\}$ to each of the nodes. If two nodes share an edge and are assigned the same color, then the nodes have a \textit{conflict} on this edge.

Our process consists of discrete time steps (\textit{states}), where we have a current coloring of the graph in every state. When a node $v$ is currently colored $c_1$, but there exists a color $c_2$ such that the neighborhood of $v$ contains strictly less nodes colored $c_2$ than nodes colored $c_1$, then the node is \textit{switchable} (since the node could reduce its number of conflicts by changing its color). The process of $v$ changing its color is \textit{switching}. Nodes always make locally optimal solutions, that is, they switch to the color which is least frequent in their neighborhood. In case of multiple optimal colors, related work on majority processes considers different tie-breaking rules. However, our constructions ensure that a tie can never occur, and thus our bounds hold for any tie-breaking strategy.

The \textit{minority process} is a sequence of steps, where each step is described by a set of nodes that switch. Note that we only consider valid steps, where every chosen node is switchable.

A state is \textit{stable} when no node in the graph is switchable; a system \textit{stabilizes} if it reaches a stable state. \textit{Stabilization time} is the number of steps until the process stabilizes. Note that in case of model E, papers studying majority processes often use a different definition of stabilization, based on periodicity. However, our constructions ensure that the process always ends in a stable state, thus for the graphs in the paper, the two definitions of stabilization are equivalent.

In our examples, we will consider the case of having only two available colors, black and white. However, as discussed in Section \ref{sec:tools}, our lower bounds are easy to generalize to any number of colors.

The restriction to two colors allows us to introduce some helpful terminology. Consider a node $v$ at a given state of the process. If $v$ has $v_s$ neighbors with the same color as $v$, and $v_o$ neighbors with the opposite color, the number $v_o-v_s$ is called the \textit{balance} of $v$. Note that if one of the neighbors of $v$ switches, then the balance of $v$ either increases or decreases by 2 (which shows that the parity of the balance of $v$ can never change). The definition also implies that $v$ is switchable if and only if its balance is negative. Switching $v$ changes the sign of its balance.

\subsection{General tools in the constructions} \label{sec:tools}

\subparagraph*{Groups.} We use the notion \textit{group} to refer to a set of nodes that have the same initial color and the exact same set of neighbors (hence, groups are independent sets). Groups are, in fact, only a tool to consider certain nodesets together as one entity for simpler presentation. They will be shown as only one node with double borders in the figures, with the size of the group indicated in brackets.

In the adversarial case, we will only consider sequences that switch groups together (i.e consecutively in any order). In the benevolent case, groups will be switched together in the sense that if a node in the group switches, then all other nodes in the group will also switch before any neighbor of the group becomes switchable; this property is enforced by the graph construction.

The more complicated definition in the benevolent case is due to the fact that we have to consider every possible sequence that the player can choose. Technically, in some sequences, a group might not be switched consecutively (it might be interrupted by switches in other, distant parts of the graph), but the outcome will still be equivalent to switching them consecutively.

\subparagraph*{Fixed nodes.} Assume we already have a graph $G$ on $n$ nodes, with maximum degree $\Delta$. Now let us add two more set of nodes $F_w$, $F_b$ to the graph such that $|F_w|=|F_b|=n+1$, and $v_w$ and $v_b$ are connected for all $v_w \in F_w$, $v_b \in F_b$. Let the color of $F_w$ and $F_b$ initially be white and black, respectively. The nodes in $F_w$ and $F_b$ will be referred to as \textit{fixed nodes}, and we will connect them to some of the nodes in our original graph. Note that these fixed nodes already have $n+1$ neighbors of the opposite color, and can never have more neighbors of the same color (as they can have at most $n$ neighbors $G$), so their color is indeed fixed and they can never switch.

Such fixed nodes are widely used in our construction; we can allow any node in $G$ to have up to $\Delta+1$ fixed neighbors of either color (having more fixed node neighbors does not affect the behavior of the node anymore). The introduction of fixed nodes increases the graph size only by a constant factor (to $3n+2$), so all lower bounds expressed as a function of $n$ will still be of the same magnitude as a function of $3n+2$. Therefore, for ease of presentation, we still use $n$ to denote the number of nodes in the graph \textit{without} the extra fixed nodes, and express our bounds as a function of $n$.

Fixed node neighbors are denoted by squares in the figures, with the multiplicity written beside the node (if more than 1). Note that for convenience, we draw separate squares for different nodes, even though the corresponding fixed node sets might overlap. This is because fix node connections are thought of as a ``property'' of the node, introducing an offset into its initial balance.

\subparagraph*{Generalization to more colors.} While the paper discusses the case of two colors, a simple idea allows a generalization to any constant number of colors $k$. Assume we have a construction $G$ on $n$ nodes, showing a lower bound on stabilization time with two colors; we can simply add sets of nodes $F_3$, $F_4$, ..., $F_k$ of size $\Delta+1$ such that they form a complete multipartite graph, and connect all these new nodes to all nodes in $G$. Let us color the nodes in $F_i$ with color $i$.

None of the original nodes in $G$ will ever assume any of the colors $3$, $4$, ..., $k$, since they always have $\Delta+1$ neighbors of these colors, while they have strictly less (at most $\Delta$) neighbors of colors $1$ and $2$. Nodes in $F_i$ will never have any incentive to switch, since they have no conflicts at all. Thus the process will behave as if the graph only consisted of $G$ with colors 1 and 2. As the new nodes only increase the graph size by a constant factor, we receive an example with the same magnitude of running time, but with $k$ colors.

With the same technique, our lower bound of $\Omega(n^{3/2})$ can also be generalized to the case of up to $\Theta(\sqrt{n})$ colors; details of this are discussed in Appendix \ref{App:A}.

\section{Sequential adversarial model} \label{sec:seqadv}

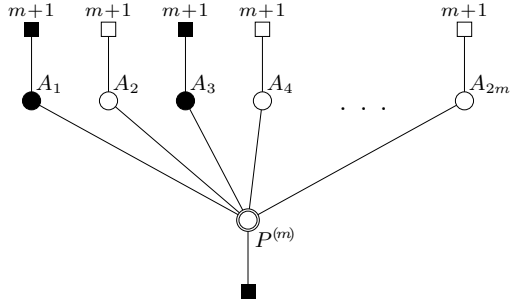
\begin{figure}
\centering
\captionsetup{justification=centering}
	\scalebox{.9}{\begin{tikzpicture}
	
	\draw (90pt,0pt) -- (0pt,50pt);
	\draw (90pt,0pt) -- (32pt,50pt);
	\draw (90pt,0pt) -- (64pt,50pt);
	\draw (90pt,0pt) -- (96pt,50pt);
	\draw (90pt,0pt) -- (180pt,50pt);
	
	\draw (90pt,0pt) -- (90pt,-30pt);
	\draw (0pt,50pt) -- (0pt,80pt);
	\draw (32pt,50pt) -- (32pt,80pt);
	\draw (64pt,50pt) -- (64pt,80pt);
	\draw (96pt,50pt) -- (96pt,80pt);
	\draw (180pt,50pt) -- (180pt,80pt);
	
	\draw[black, fill=white] (90pt,0pt) circle (1ex);
	\draw[black, fill=white] (90pt,0pt) circle (0.8ex);
		
	\draw[fill=black] (0pt,50pt) circle (0.8ex);
	\draw[black, fill=white] (32pt,50pt) circle (0.8ex);
	\draw[fill=black] (64pt,50pt) circle (0.8ex);
	\draw[black, fill=white] (96pt,50pt) circle (0.8ex);
	\node[anchor=north] at (138pt,51pt) {\large . . .};
	\draw[black, fill=white] (180pt,50pt) circle (0.8ex);
	
	\fill[black] (87pt,-33pt) rectangle (93pt,-27pt);
	\fill[black] (-3pt,77pt) rectangle (3pt,83pt);
	\draw[black, fill=white] (29pt,77pt) rectangle (35pt,83pt);
	\fill[black] (61pt,77pt) rectangle (67pt,83pt);
	\draw[black, fill=white] (93pt,77pt) rectangle (99pt,83pt);
	\draw[black, fill=white] (177pt,77pt) rectangle (183pt,83pt);
	
	\node[anchor=south] at (0pt,81pt) {\scriptsize $m\!+\!1$};
	\node[anchor=south] at (32pt,81pt) {\scriptsize $m\!+\!1$};
	\node[anchor=south] at (64pt,81pt) {\scriptsize $m\!+\!1$};
	\node[anchor=south] at (96pt,81pt) {\scriptsize $m\!+\!1$};
	\node[anchor=south] at (180pt,81pt) {\scriptsize $m\!+\!1$};
	
	\node[anchor=west] at (-2pt,57pt) {\footnotesize $A_1$};
	\node[anchor=west] at (30pt,57pt) {\footnotesize $A_2$};
	\node[anchor=west] at (62pt,57pt) {\footnotesize $A_3$};
	\node[anchor=west] at (94pt,57pt) {\footnotesize $A_4$};
	\node[anchor=west] at (178pt,57pt) {\footnotesize $A_{2m}$};
	\node[anchor=west] at (89pt,-7pt) {\footnotesize $P$\footnotesize$^{(\!m\!)}$};
	
\end{tikzpicture}}
	\caption{Construction with an adversarial sequence of $\Theta(n^2)$ switches}
	\label{fig:adverse}
\end{figure}

We first present a simple example where model A takes $\Omega(n^2)$ steps. Let us introduce a parameter $m$, the value of which will be determined later. Our construction, shown in Figure \ref{fig:adverse}, consist of a group $P$ of size $m$, initially colored white, and $2m$ distinct nodes $A_1$, $A_2$, ..., $A_{2m}$, such that $A_i$ is initially colored black for odd values of $i$ and white for even $i$. Let us connect all nodes $A_i$ to $P$, and add one more fixed black node that is connected only to $P$. Finally, let us connect each $A_i$ to $m+1$ fixed nodes of the same color as $A_i$. Recall that although the figure shows multiple squares, there are in fact only $n+1$ fixed black and $n+1$ fixed white nodes in the graph altogether.

In this graph, $P$ has a balance of 1 initially, while black $A_i$ have a balance of $-1$ and white $A_i$ have a balance of $-(2m+1)$. Note that even after execution begins, until $A_i$ is switched for the first time, it will have $m+1$ fixed neighbors of the same color and at most $m$ neighbors of the opposite color (depending on the current color of $P$), and thus a negative balance. Therefore, each $A_i$ is switchable anytime if it has not been switched before.

Consider the following sequence of adversarial moves in this graph: the player first decides to switch $A_1$, then $P$, then $A_2$, then $P$ again, then $A_3$, $P$, ..., $A_{2m}$, and finally $P$ again. As each $A_i$ is used only once, they are clearly all switchable. As for $P$, its balance first changes from 1 to $-1$, when changing $A_1$ to white, but increases back to $1$ when we switch $P$ itself. Then it changes to $-1$ once again after changing $A_2$, so it is switchable again, and so on: each time we switch an $A_i$, we change it to the same color that $P$ currently has, decreasing $P$'s balance to $-1$, which increases back to 1 again as we switch $P$. Therefore, this strategy is indeed a sequence of valid switches.

Since $P$ contains $m$ nodes and is switched $2m$ times in this sequence, this alone contributes to $2m^2$ switches. Altogether, we have $3m$ nodes in the graph, allowing us a choice of $m=\frac{n}{3}$ (recall that fixed nodes are ignored when counting the nodes in the graph). This gives us a sequence with at least $\frac{2}{9}n^2$ steps.

\begin{theorem}
There exists a graph construction with $\Omega(n^2)$ stabilization time in model A.
\end{theorem}

\section{Construction for benevolent models} \label{sec:seqben}

This section presents a graph construction with $\Omega(n^{3/2})$ stabilization time in benevolent models. Note that it is much more involved to find an example where benevolent models take $\omega(n)$ steps, since in such a construction, we have to ensure that any possible sequence lasts for a long time. In order to have an easy-to-analyze construction, our graph will, at any point in time, contain only one, or a small given set of nodes that are switchable, and switching this or these nodes enables the next such set of nodes (i.e., makes them switchable). This way, the switchable point ``propagates'' through the graph, and the benevolent player has no other valid move than to follow this path of propagation that has been designed into the graph.

The general idea behind the construction is to have a linearly long chain of nodes which is propagated through multiple times. After each such round, the propagation enters a different branch of further nodes; this branch resets the chain for the following round, and then also triggers the following round of propagation (as outlined later in Figure \ref{fig:system}).

Due to the complexity of the construction, we do not describe it directly; instead, we define smaller functional elements (\textit{gadgets}) that execute a certain task. We then use these gadgets as building blocks to put our example graph together. This section outlines the tasks and main properties of the gadgets; a detailed description and analysis of each gadget can be found in Appendix \ref{App:A}. While the concrete gadget designs are specific to minority processes, they are built on general ideas and techniques for benevolent models; as such, we hope they may inspire similar solutions in the analysis of related processes or cellular automata.

When describing a gadget, the edges connecting the gadget to other nodes in the graphs are drawn as dashed lines in the figures, with the external node usually denoted by $v$ (possibly with some subscript). Although our graph is undirected, we often refer to such edges as \textit{input} or \textit{output} edges of the gadget, and also show this direction in the figures. This will refer to the role that the external node plays in the functionality of the gadget. That is, whenever the gadget is used in our constructions, it is triggered by (some of) its input nodes switching, and upon completing its task, the gadget makes (some of) its output nodes switchable.

Naturally, as in the entire graph, the role of the two colors is always interchangeable within the gadgets. Therefore, we only present each such gadget in one color variant.

Due to the complexity of the construction, we have also verified its correctness through implementing the process. A discussion of these simulations is available in Appendix \ref{App:C}.

\subparagraph*{Simple relay.}
As our most basic tool to propagate the only possible point of switching, we use the \textit{simple relay} gadget shown in Figure \ref{fig:relay}. A simple relay only consist of a base node $B$, connected to a fixed node of the same color.

\begin{figure}
\centering
\captionsetup{justification=centering}
\begin{subfigure}[b]{0.157\textwidth}
	\raisebox{-7pt}{
		\scalebox{.95}{\begin{tikzpicture}
	\draw[dashed, arrows=-stealth] (-5pt,0pt) -- (11.5pt,0pt);
	\draw[dashed, arrows=-stealth] (15pt,0pt) -- (32pt,0pt);
	\draw (15pt,20pt) -- (15pt,0pt);
	\draw[black, fill=white] (15pt,0pt) circle (0.8ex);
	\draw[black, fill=white] (12pt,17pt) rectangle (18pt,23pt);
	
	\draw[fill=black] (-5pt,0pt) circle (0.7ex);
	\draw[fill=black] (35pt,0pt) circle (0.7ex);
	
	\node[anchor=west] at (13pt,-8pt) {\footnotesize $B$};
	\node[anchor=east] at (-1pt,-8pt) {\footnotesize $v_L$};
	\node[anchor=west] at (34pt,-8pt) {\footnotesize $v_R$};
\end{tikzpicture}}
	}
	\caption{}
	\label{fig:relay}
\end{subfigure}
\hspace{0.06\textwidth}
\begin{subfigure}[b]{0.405\textwidth}
  \raisebox{10pt}{ 
		\scalebox{.95}{\begin{tikzpicture}
	\draw (0pt,0pt) -- (15pt,0pt);
	\draw (15pt,0pt) -- (30pt,0pt);
	\draw (15pt,20pt) -- (15pt,0pt);
	\draw[black, fill=white] (15pt,0pt) circle (0.8ex);
	\draw[black, fill=white] (12pt,17pt) rectangle (18pt,23pt);
	\draw[fill=black] (0pt,0pt) circle (0.6ex);
	\draw[fill=black] (30pt,0pt) circle (0.6ex);
	
	\draw[very thick, arrows=-stealth] (32pt,14pt) -- (48pt,14pt);
	
	\draw (50pt,0pt) -- (65pt,0pt);
	\draw (65pt,0pt) -- (80pt,0pt);
	\draw (65pt,20pt) -- (65pt,0pt);
	\draw[black, fill=white] (65pt,0pt) circle (0.8ex);
	\draw[black, fill=white] (62pt,17pt) rectangle (68pt,23pt);
	\draw[black, fill=white] (50pt,0pt) circle (0.6ex);
	\draw[fill=black] (80pt,0pt) circle (0.6ex);
	
	\draw[very thick, arrows=-stealth] (82pt,14pt) -- (98pt,14pt);
	
	\draw (100pt,0pt) -- (115pt,0pt);
	\draw (115pt,0pt) -- (130pt,0pt);
	\draw (115pt,20pt) -- (115pt,0pt);
	\draw[fill=black] (115pt,0pt) circle (0.8ex);
	\draw[black, fill=white] (112pt,17pt) rectangle (118pt,23pt);
	\draw[black, fill=white] (100pt,0pt) circle (0.6ex);
	\draw[fill=black] (130pt,0pt) circle (0.6ex);
	
	\draw[very thick, arrows=-stealth] (132pt,14pt) -- (148pt,14pt);
	
	\draw (150pt,0pt) -- (165pt,0pt);
	\draw (165pt,0pt) -- (180pt,0pt);
	\draw (165pt,20pt) -- (165pt,0pt);
	\draw[fill=black] (165pt,0pt) circle (0.8ex);
	\draw[black, fill=white] (162pt,17pt) rectangle (168pt,23pt);
	\draw[black, fill=white] (150pt,0pt) circle (0.6ex);
	\draw[black, fill=white] (180pt,0pt) circle (0.6ex);
\end{tikzpicture}}
	}
	\caption{}
	\label{fig:relay_phases}
\end{subfigure}
\hspace{0.06\textwidth}
\begin{subfigure}[b]{0.27\textwidth}
	\raisebox{10pt}{
		\scalebox{.95}{\begin{tikzpicture}
	\draw (0pt,0pt) -- (15pt,0pt);
	\draw (15pt,0pt) -- (45pt,0pt);
	\draw (15pt,20pt) -- (15pt,0pt);
	\draw[black, fill=white] (15pt,0pt) circle (0.8ex);
	\draw[black, fill=white] (12pt,17pt) rectangle (18pt,23pt);
	
	\draw (45pt,0pt) -- (75pt,0pt);
	\draw (45pt,20pt) -- (45pt,0pt);
	\draw[fill=black] (45pt,0pt) circle (0.8ex);
	\draw[fill=black] (42pt,17pt) rectangle (48pt,23pt);
	
	\draw (75pt,0pt) -- (105pt,0pt);
	\draw (75pt,20pt) -- (75pt,0pt);
	\draw[black, fill=white] (75pt,0pt) circle (0.8ex);
	\draw[black, fill=white] (72pt,17pt) rectangle (78pt,23pt);
	
	\draw (105pt,0pt) -- (120pt,0pt);
	\draw (105pt,20pt) -- (105pt,0pt);
	\draw[fill=black] (105pt,0pt) circle (0.8ex);
	\draw[fill=black] (102pt,17pt) rectangle (108pt,23pt);

\end{tikzpicture}}
	}
	\caption{}
	\label{fig:relay_chain}
\end{subfigure}
\caption{Simple relay gadget (a), the steps of its operation (b), and a chain of relays (c)}\label{fig:allRelay}
\end{figure}

Besides the fixed node, $B$ also has a left and a right neighbor ($v_L$ and $v_R$) outside of the gadget, both of which initially have the opposite color as $B$. This way, until neither of the two switch, $B$ has positive balance and cannot switch either. However, as soon as $v_L$ switches to the color of $B$, $B$ becomes switchable, and as $B$ switches, this can propagate the point of change to its other neighbor $v_R$ (Figure \ref{fig:relay_phases}).

Note that connecting alternating-colored relays into a chain already gives a simple example of linear stabilization time (see Figure \ref{fig:relay_chain}). If the leftmost (white) relay's base node is connected to a fixed white node, then the only available sequence of moves is to switch the base nodes in the relays one by one from left to right, resulting in a sequence of $n$ steps.

Through the concept of input and output nodes, relays essentially allow us to connect other, more sophisticated gadgets in our constructions. If some gadget has an output node $v_1$ and another gadget has an input node $v_2$, we can add a chain of relays between $v_1$ and $v_2$, ensuring that once $v_1$ switches, it will be followed by $v_2$ eventually. Because of this role, simple relays are not shown explicitly in our final overview figure of the construction, but only represented by arrows, indicating the direction of propagation between more complex gadgets.

\subparagraph*{Rechargeable relay.}
A more sophisticated version of a relay is the \textit{rechargeable relay} shown in Figure \ref{fig:rechargeable}. In such a relay, node $B$ is extended by an upper node $U$, a control group $C$ of size 2, and two recharge nodes $R_1$, $R_2$, the role of which are interchangeable. Besides $v_L$ and $v_R$, the nodes $R_1$ and $R_2$ also have edges to some external nodes. It is always ensured that the initial balance of $R_1$ and $R_2$ from these upper neighbors (that is, with $C$ ignored) is exactly 3.

As in case of a simple relay, if $v_L$ switches, then $B$ itself can switch, followed by $v_R$. Now assume that in this ``used'' phase of the relay, some outside circumstance changes 3 neighbors of node $R_2$ from black to white, and thus its balance changes from the current value of 5 to $-1$ (the relay is \textit{recharged}). Then $R_2$ can switch to black, making $C$ and in turn $U$ switch, too. Finally, assume that some other outside circumstance then changes the balance of $R_2$ from 5 to $-1$ again (known as \textit{resetting} the relay); then $R_2$ will switch back to white (with a new balance of 1), and we end up in the initial state of a rechargeable relay of the opposite color. The steps of the process are shown in Figure \ref{fig:recharge_phases}.

\begin{figure}
\centering
\captionsetup{justification=centering}
\begin{subfigure}[b]{0.157\textwidth}
	\raisebox{-7pt}{
		\scalebox{.95}{\begin{tikzpicture}
	\draw[dashed, arrows=-stealth] (-5pt,0pt) -- (11.5pt,0pt);
	\draw[dashed, arrows=-stealth] (15pt,0pt) -- (32pt,0pt);
	\draw (15pt,14pt) -- (15pt,0pt);
	\draw (15pt,14pt) -- (15pt,29pt);
	\draw (2pt,40pt) -- (15pt,29pt);
	\draw (28pt,40pt) -- (15pt,29pt);
	\draw[dashed, arrows=stealth-] (-1pt,40pt) -- (-13pt,42pt);
	\draw[dashed, arrows=stealth-] (0pt,42pt) -- (-9pt,51pt);
	\draw[dashed, arrows=stealth-] (2pt,43pt) -- (0pt,55pt);
	\draw[dashed, arrows=stealth-] (31pt,40pt) -- (43pt,42pt);
	\draw[dashed, arrows=stealth-] (30pt,42pt) -- (39pt,51pt);
	\draw[dashed, arrows=stealth-] (28pt,43pt) -- (30pt,55pt);
	\draw[black, fill=white] (15pt,0pt) circle (0.7ex);
	\draw[black, fill=white] (15pt,14pt) circle (0.7ex);
	\draw[black, fill=white] (15pt,29pt) circle (1ex);
	\draw[fill=black] (15pt,29pt) circle (0.8ex);
	
	\draw[fill=black] (2pt,40pt) circle (0.7ex);
	\draw[black, fill=white] (28pt,40pt) circle (0.7ex);
	\draw[fill=black] (-5pt,0pt) circle (0.7ex);
	\draw[fill=black] (35pt,0pt) circle (0.7ex);
	
	\node[anchor=west] at (13pt,-8pt) {\scriptsize $B$};
	\node[anchor=west] at (15pt,9pt) {\scriptsize $U$};
	\node[anchor=west] at (15pt,23pt) {\scriptsize $C$\tiny$^{(2)}$};
	\node[anchor=east] at (5pt,34pt) {\scriptsize $R_1$};
	\node[anchor=west] at (27pt,34pt) {\scriptsize $R_2$};
	\node[anchor=east] at (-1pt,-8pt) {\scriptsize $v_L$};
	\node[anchor=west] at (33pt,-8pt) {\scriptsize $v_R$};
\end{tikzpicture}}
	}
	\caption{}
	\label{fig:rechargeable}
\end{subfigure}
\hspace{0.07\textwidth}
\begin{subfigure}[b]{0.72\textwidth}
  \raisebox{10pt}{ 
		\scalebox{.95}{\begin{tikzpicture}
	\draw (-5pt,0pt) -- (15pt,0pt);
	\draw (15pt,0pt) -- (35pt,0pt);
	\draw (15pt,14pt) -- (15pt,0pt);
	\draw (15pt,14pt) -- (15pt,29pt);
	\draw (2pt,40pt) -- (15pt,29pt);
	\draw (28pt,40pt) -- (15pt,29pt);
	\draw (2pt,40pt) -- (-13pt,42pt);
	\draw (2pt,40pt) -- (-10pt,52pt);
	\draw (2pt,40pt) -- (0pt,55pt);
	\draw (28pt,40pt) -- (43pt,42pt);
	\draw (28pt,40pt) -- (40pt,52pt);
	\draw (28pt,40pt) -- (30pt,55pt);
	\draw[fill=black] (15pt,0pt) circle (0.7ex);
	\draw[black, fill=white] (15pt,14pt) circle (0.7ex);
	\draw[black, fill=white] (15pt,29pt) circle (1ex);
	\draw[fill=black] (15pt,29pt) circle (0.8ex);
	\node[anchor=west] at (15pt,23pt) {\tiny $(2)$};
	\draw[fill=black] (2pt,40pt) circle (0.7ex);
	\draw[black, fill=white] (28pt,40pt) circle (0.7ex);
	\draw[black, fill=white] (-5pt,0pt) circle (0.7ex);
	\draw[black, fill=white] (35pt,0pt) circle (0.7ex);
	
	\draw (28pt,57pt) -- (28pt,60pt) -- (46pt,60pt) -- (46pt,57pt);
	\draw (37pt,60pt) -- (37pt,63pt);
	\node[anchor=south] at (37pt,72pt) {\scriptsize 3 switches};
	\node[anchor=south] at (37pt,62pt) {\scriptsize to white};
	
	\draw[very thick, arrows=-stealth] (41pt,20pt) -- (59pt,20pt);
	
	\draw (65pt,0pt) -- (85pt,0pt);
	\draw (85pt,0pt) -- (105pt,0pt);
	\draw (85pt,14pt) -- (85pt,0pt);
	\draw (85pt,14pt) -- (85pt,29pt);
	\draw (72pt,40pt) -- (85pt,29pt);
	\draw (98pt,40pt) -- (85pt,29pt);
	\draw (72pt,40pt) -- (57pt,42pt);
	\draw (72pt,40pt) -- (60pt,52pt);
	\draw (72pt,40pt) -- (70pt,55pt);
	\draw (98pt,40pt) -- (113pt,42pt);
	\draw (98pt,40pt) -- (110pt,52pt);
	\draw (98pt,40pt) -- (100pt,55pt);
	\draw[fill=black] (85pt,0pt) circle (0.7ex);
	\draw[black, fill=white] (85pt,14pt) circle (0.7ex);
	\draw[black, fill=white] (85pt,29pt) circle (1ex);
	\draw[fill=black] (85pt,29pt) circle (0.8ex);
	\node[anchor=west] at (85pt,23pt) {\tiny $(2)$};
	\draw[fill=black] (72pt,40pt) circle (0.7ex);
	\draw[fill=black] (98pt,40pt) circle (0.7ex);
	\draw[black, fill=white] (65pt,0pt) circle (0.7ex);
	\draw[black, fill=white] (105pt,0pt) circle (0.7ex);
	
	\draw[very thick, arrows=-stealth] (111pt,20pt) -- (129pt,20pt);
	
	\draw (135pt,0pt) -- (155pt,0pt);
	\draw (155pt,0pt) -- (175pt,0pt);
	\draw (155pt,14pt) -- (155pt,0pt);
	\draw (155pt,14pt) -- (155pt,26pt);
	\draw (142pt,40pt) -- (155pt,26pt);
	\draw (168pt,40pt) -- (155pt,26pt);
	\draw (142pt,40pt) -- (127pt,42pt);
	\draw (142pt,40pt) -- (130pt,52pt);
	\draw (142pt,40pt) -- (140pt,55pt);
	\draw (168pt,40pt) -- (183pt,42pt);
	\draw (168pt,40pt) -- (180pt,52pt);
	\draw (168pt,40pt) -- (170pt,55pt);
	\draw[fill=black] (155pt,0pt) circle (0.7ex);
	\draw[black, fill=white] (155pt,14pt) circle (0.7ex);
	\draw[black, fill=white] (155pt,29pt) circle (1ex);
	\draw[black, fill=white] (155pt,29pt) circle (0.8ex);
	\node[anchor=west] at (155pt,23pt) {\tiny $(2)$};
	\draw[fill=black] (142pt,40pt) circle (0.7ex);
	\draw[fill=black] (168pt,40pt) circle (0.7ex);
	\draw[black, fill=white] (135pt,0pt) circle (0.7ex);
	\draw[black, fill=white] (175pt,0pt) circle (0.7ex);
	
	\draw[very thick, arrows=-stealth] (181pt,20pt) -- (199pt,20pt);
	
	\draw (205pt,0pt) -- (225pt,0pt);
	\draw (225pt,0pt) -- (245pt,0pt);
	\draw (225pt,14pt) -- (225pt,0pt);
	\draw (225pt,14pt) -- (225pt,26pt);
	\draw (212pt,40pt) -- (225pt,26pt);
	\draw (238pt,40pt) -- (225pt,26pt);
	\draw (212pt,40pt) -- (197pt,42pt);
	\draw (212pt,40pt) -- (200pt,52pt);
	\draw (212pt,40pt) -- (210pt,55pt);
	\draw (238pt,40pt) -- (253pt,42pt);
	\draw (238pt,40pt) -- (250pt,52pt);
	\draw (238pt,40pt) -- (240pt,55pt);
	\draw[fill=black] (225pt,0pt) circle (0.7ex);
	\draw[fill=black] (225pt,14pt) circle (0.7ex);
	\draw[black, fill=white] (225pt,29pt) circle (1ex);
	\draw[black, fill=white] (225pt,29pt) circle (0.8ex);
	\node[anchor=west] at (225pt,23pt) {\tiny $(2)$};
	\draw[fill=black] (212pt,40pt) circle (0.7ex);
	\draw[fill=black] (238pt,40pt) circle (0.7ex);
	\draw[black, fill=white] (205pt,0pt) circle (0.7ex);
	\draw[black, fill=white] (245pt,0pt) circle (0.7ex);
	
	\draw (238pt,57pt) -- (238pt,60pt) -- (256pt,60pt) -- (256pt,57pt);
	\draw (247pt,60pt) -- (247pt,63pt);
	\node[anchor=south] at (247pt,72pt) {\scriptsize 3 switches};
	\node[anchor=south] at (247pt,62pt) {\scriptsize to black};
	
	\draw[very thick, arrows=-stealth] (251pt,20pt) -- (269pt,20pt);
	
	\draw (275pt,0pt) -- (295pt,0pt);
	\draw (295pt,0pt) -- (315pt,0pt);
	\draw (295pt,14pt) -- (295pt,0pt);
	\draw (295pt,14pt) -- (295pt,26pt);
	\draw (282pt,40pt) -- (295pt,26pt);
	\draw (308pt,40pt) -- (295pt,26pt);
	\draw (282pt,40pt) -- (267pt,42pt);
	\draw (282pt,40pt) -- (270pt,52pt);
	\draw (282pt,40pt) -- (280pt,55pt);
	\draw (308pt,40pt) -- (323pt,42pt);
	\draw (308pt,40pt) -- (320pt,52pt);
	\draw (308pt,40pt) -- (310pt,55pt);
	\draw[fill=black] (295pt,0pt) circle (0.7ex);
	\draw[fill=black] (295pt,14pt) circle (0.7ex);
	\draw[black, fill=white] (295pt,29pt) circle (1ex);
	\draw[black, fill=white] (295pt,29pt) circle (0.8ex);
	\node[anchor=west] at (295pt,23pt) {\tiny $(2)$};
	\draw[fill=black] (282pt,40pt) circle (0.7ex);
	\draw[black, fill=white] (308pt,40pt) circle (0.7ex);
	\draw[black, fill=white] (275pt,0pt) circle (0.7ex);
	\draw[black, fill=white] (315pt,0pt) circle (0.7ex);
	
\end{tikzpicture}}
	}
	\caption{}
	\label{fig:recharge_phases}
\end{subfigure}
\caption{Rechargeable relay gadget (a) and the steps of its operation (b)}\label{fig:allRecRelay}
\end{figure}

This is exactly the essence of this gadget: it is a relay which can be used the same way multiple times. Connecting such gadgets into a chain in the same fashion as Figure \ref{fig:relay_chain}, we get a chain that can propagate the point of change not only once, but multiple times if ``recharged'' through their upper connections between two such propagations.

\subparagraph*{Recharging system.}
The rechargeable relay suggests that it is useful to have a tool to ``recharge'' some nodes, i.e. to decrease their balance by switching some of their neighbors to the color they currently have. To execute this task efficiently on many nodes, we present a \textit{recharging system}.

For the first version of this gadget, assume a setting where there is a set $X$ of $m$ black nodes, and we want to decrease the balance of each of these nodes by $2$ (i.e., change exactly one white neighbor of each of them to black). A \textit{basic recharging system}, shown in Figure \ref{fig:recharger},
can execute this task while using only $O(\sqrt{m})$ nodes.
The gadget is organized into 3 levels: a single node $U$ in the upper level, a group $M$ of $\sqrt{m}+1$ nodes in the middle level, and $\sqrt{m}$ distinct nodes $L_i$ in the lower level. Each lower level node is connected to $\sqrt{m}$ different nodes in $X$, thus exactly covering the nodes of $X$.

\begin{figure}
\centering
\captionsetup{justification=centering}
%\hspace{-0.08\textwidth}
\begin{subfigure}[b]{0.287\textwidth}
	\raisebox{0pt}{
		\begin{tikzpicture}

	\draw (0pt,0pt) -- (25pt,25pt);
	\draw (10pt,0pt) -- (25pt,25pt);
	\draw (20pt,0pt) -- (25pt,25pt);
	\draw (50pt,0pt) -- (25pt,25pt);
	
	\draw[dashed, arrows=-stealth] (0pt,0pt) -- (-12pt,-20pt);
	\draw[dashed, arrows=-stealth] (0pt,0pt) -- (-8pt,-20pt);
	\draw[dashed, arrows=-stealth] (0pt,0pt) -- (-4pt,-20pt);
	
	\draw[dashed, arrows=-stealth] (10pt,0pt) -- (6pt,-20pt);
	\draw[dashed, arrows=-stealth] (10pt,0pt) -- (10pt,-20pt);
	\draw[dashed, arrows=-stealth] (10pt,0pt) -- (14pt,-20pt);
	
	\draw[fill=black] (-13pt,-22pt) circle (0.3ex);
	\draw[fill=black] (-8.5pt,-22pt) circle (0.3ex);
	\draw[fill=black] (-4pt,-22pt) circle (0.3ex);
	
	\draw[fill=black] (5.5pt,-22pt) circle (0.3ex);
	\draw[fill=black] (10pt,-22pt) circle (0.3ex);
	\draw[fill=black] (14.5pt,-22pt) circle (0.3ex);
	
	\draw[dashed] (20pt,0pt) -- (20pt,-11pt);
	\draw[dashed] (20pt,0pt) -- (22pt,-10pt);
	\draw[dashed] (20pt,0pt) -- (24pt,-9pt);
	
	\draw[dashed] (50pt,0pt) -- (50pt,-9pt);
	\draw[dashed] (50pt,0pt) -- (52pt,-11pt);
	\draw[dashed] (50pt,0pt) -- (54pt,-9pt);
	
	\draw (-15pt,-26pt) -- (-15pt,-28pt) -- (-2pt,-28pt) -- (-2pt,-26pt);
	\draw (-8.5pt,-28pt) -- (-8.5pt,-30pt);
	\node at (-8.5pt,-33pt) {\tiny \textit{$\sqrt{m}$ nodes}};
	
	\draw (-15pt,-42pt) -- (-15pt,-44pt) -- (59pt,-44pt) -- (59pt,-42pt);
	\draw (22pt,-44pt) -- (22pt,-46pt);
	\node at (22pt,-51pt) {\tiny Set $X$ of $m$ nodes};
	
	\draw (25pt,25pt) -- (25pt,50pt);
	
	\draw[dashed, arrows=stealth-] (28pt,50pt) -- (60pt,50pt);
	\draw (-5pt,50pt) -- (25pt,50pt);
	
	\draw (-5pt,25pt) -- (25pt,25pt);

	\draw[black, fill=white] (0pt,0pt) circle (0.6ex);
	\draw[black, fill=white] (10pt,0pt) circle (0.6ex);
	\draw[black, fill=white] (20pt,0pt) circle (0.6ex);
	\node[anchor=west] at (26pt,-2pt) {\small \textit{...}};
	\draw[black, fill=white] (50pt,0pt) circle (0.6ex);
	
	\draw (61pt,4pt) -- (63pt,4pt) -- (63pt,-4pt) -- (61pt,-4pt);
	\draw (63pt,0pt) -- (65pt,0pt);
	
	\node[anchor=west] at (64pt,0pt) {\scriptsize$L_i$ \tiny($\sqrt{m}$ nodes)};
	
	\draw[black, fill=white] (25pt,25pt) circle (0.9ex);
	\draw[fill=black] (25pt,25pt) circle (0.7ex);
	
	\node[anchor=west] at (29pt,26pt) {\scriptsize$M$\tiny$^{(\sqrt{m}+1)}$};
	
	\draw[black, fill=white] (25pt,50pt) circle (0.6ex);
	
	\node[anchor=west] at (24pt,57pt) {\scriptsize $U$};
	
	\draw[fill=black] (60pt,50pt) circle (0.45ex);
	
	\node[anchor=west] at (60pt,53pt) {\scriptsize $v$};
	
	\draw[black, fill=white] (-7pt,52pt) rectangle (-3pt,48pt);
	
	\node[anchor=east] at (-7pt,50pt) {\tiny $\sqrt{m}+1$};
	
	\fill[black] (-7pt,27pt) rectangle (-3pt,23pt);
	
	\node[anchor=east] at (-7pt,25pt) {\tiny $\sqrt{m}$};
	
\end{tikzpicture}
	}
	\caption{}
	\label{fig:recharger}
\end{subfigure}
\hspace{0.19\textwidth}
\begin{subfigure}[b]{0.295\textwidth}
  \raisebox{6.5pt}{ 
		\begin{tikzpicture}

	\draw (0pt,0pt) -- (25pt,25pt);
	\draw (10pt,0pt) -- (25pt,25pt);
	\draw (20pt,0pt) -- (25pt,25pt);
	\draw (50pt,0pt) -- (25pt,25pt);
	
	\draw[dashed, arrows=-stealth] (0pt,0pt) -- (-1pt,-20pt);
	\draw[dashed, arrows=-stealth] (0pt,0pt) -- (4pt,-20pt);
	\draw[dashed, arrows=-stealth] (0pt,0pt) -- (10.5pt,-20pt);
	
	\draw[dashed, arrows=-stealth] (10pt,0pt) -- (6pt,-20pt);
	%\draw[dashed, arrows=-stealth] (10pt,0pt) -- (11pt,-20pt);
	\draw[dashed, arrows=-stealth] (10pt,0pt) -- (15pt,-20pt);
	
	\draw[fill=black] (-1pt,-22pt) circle (0.3ex);
	\draw[fill=black] (5pt,-22pt) circle (0.3ex);
	\draw[fill=black] (11pt,-22pt) circle (0.3ex);
	\draw[fill=black] (16pt,-22pt) circle (0.3ex);
	\draw[fill=black] (21pt,-22pt) circle (0.3ex);
	
	\draw[dashed, arrows=-stealth] (20pt,0pt) -- (17pt,-20pt);
	\draw[dashed, arrows=-stealth] (20pt,0pt) -- (21pt,-20pt);
	\draw[dashed] (20pt,0pt) -- (24pt,-9pt);
	
	\draw[dashed] (50pt,0pt) -- (50pt,-9pt);
	\draw[dashed] (50pt,0pt) -- (48pt,-11pt);
	\draw[dashed] (50pt,0pt) -- (46pt,-9pt);
	
	\draw (-2pt,-27pt) -- (-2pt,-29pt) -- (52pt,-29pt) -- (52pt,-27pt);
	\draw (25pt,-29pt) -- (25pt,-31pt);
	\node at (25pt,-37pt) {\tiny Set $X$, with a balance to be};
	\node at (25pt,-44pt) {\tiny decreased by $2 \! \cdot \! \chi$ altogether};
	
	\draw (25pt,25pt) -- (25pt,50pt);
	
	\draw[dashed, arrows=stealth-] (28pt,50pt) -- (60pt,50pt);
	\draw (-5pt,50pt) -- (25pt,50pt);
	
	\draw (-5pt,25pt) -- (25pt,25pt);

	\draw[black, fill=white] (0pt,0pt) circle (0.6ex);
	\draw[black, fill=white] (10pt,0pt) circle (0.6ex);
	\draw[black, fill=white] (20pt,0pt) circle (0.6ex);
	\node[anchor=west] at (26pt,-2pt) {\small \textit{...}};
	\draw[black, fill=white] (50pt,0pt) circle (0.6ex);
	
	\draw (61pt,4pt) -- (63pt,4pt) -- (63pt,-4pt) -- (61pt,-4pt);
	\draw (63pt,0pt) -- (65pt,0pt);
	
	\node[anchor=west] at (64pt,0pt) {\scriptsize$L_i$ \tiny($\sqrt{\chi}$ nodes)};
	
	\draw[black, fill=white] (25pt,25pt) circle (0.9ex);
	\draw[fill=black] (25pt,25pt) circle (0.7ex);
	
	\node[anchor=west] at (29pt,26pt) {\scriptsize$M$\tiny$^{(\sqrt{\chi}+1)}$};
	
	\draw[black, fill=white] (25pt,50pt) circle (0.6ex);
	
	\node[anchor=west] at (24pt,57pt) {\scriptsize $U$};
	
	\draw[fill=black] (60pt,50pt) circle (0.45ex);
	
	\node[anchor=west] at (60pt,53pt) {\scriptsize $v$};
	
	\draw[black, fill=white] (-7pt,52pt) rectangle (-3pt,48pt);
	
	\node[anchor=east] at (-7pt,50pt) {\tiny $\sqrt{\chi}+1$};
	
	\fill[black] (-7pt,27pt) rectangle (-3pt,23pt);
	
	\node[anchor=east] at (-7pt,25pt) {\tiny $\sqrt{\chi}$};
	
\end{tikzpicture}
	}
	\caption{}
	\label{fig:recharger_gen}
\end{subfigure}
\caption{Basic (a) and generalized (b) recharging system}\label{fig:allRecharger}
\end{figure}
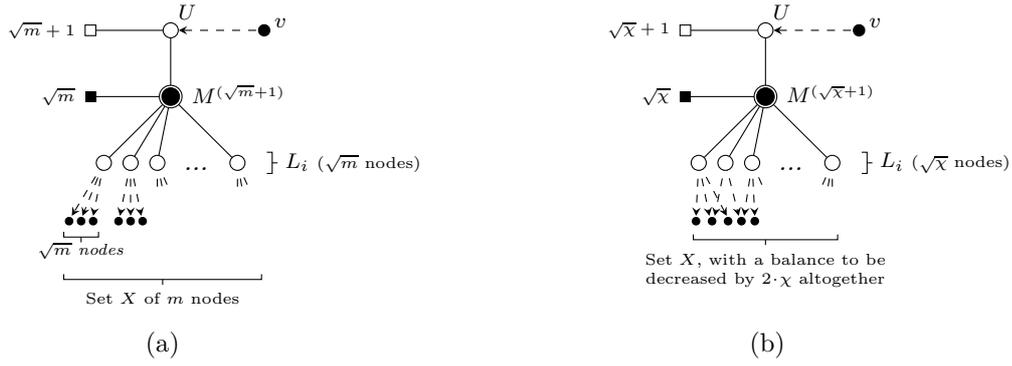

The gadget operates in a top-to-bottom fashion: once $v$ switches, $U$ turns black, followed by $M$ turning white. Once all nodes in $M$ are switched, the nodes $L_i$ all decide to switch, too.

The key idea in the design of the gadget is that each node $L_i$ has strictly more neighbors in $M$ than in $X$. This ensures that as long as $M$ is black, the nodes $L_i$ always have a positive balance, regardless of the current color of their neighbors in $X$. Therefore, no node in the gadget can ever switch before the node $U$ is triggered.

We can use this insight to create a similar gadget for a more general setting. Assume that we similarly have a set $X$ of $m$ black nodes, but instead of decreasing their balance by 2, we want to decrease the balance of each node in $X$ by some specific (possibly different) even value, denoted by $2x_1$, $2x_2$, ..., $2x_m$ (i.e., for the $j^{\text{th}}$ node in $X$, we want to change $x_j$ of its white neighbors to black). Let us denote the sum $\sum_{j=1}^{m} x_j$ of these values by $\chi$.

We can achieve this using a similar construction, shown in Figure \ref{fig:recharger_gen}. In this \textit{generalized recharging system}, we allow multiple nodes $L_i$ to be connected to the same node in $X$: if a node in $X$ has a corresponding value $2x_j$, then it has exactly $x_j$ neighbors in the lower level of the system. This ensures that once all the nodes $L_i$ switch, the new balance of each node in $X$ is exactly as desired. The number of nodes in the gadget can be minimized by placing $\sqrt{\chi}$ nodes $L_i$ in the lower level, each with $\sqrt{\chi}$ neighbors in $X$; this way, the overall number of edges going into the set $X$ from the gadget is exactly $\chi$ as required. To ensure that the neighborhood of each $L_i$ is dominated by $M$, we choose the size of group $M$ to be $\sqrt{\chi}+1$.

\subparagraph*{AND gate.}
Another ingredient we use is an \textsc{and} gate. As its name suggests, this gadget has $x$ input edges from a set of nodes $X$, and once all nodes in $X$ have switched to the same color (say, white), the gadget triggers a change in another part of the graph.

\begin{figure}
\captionsetup{justification=centering}
\minipage{0.3\textwidth}
	\vspace{22pt}
	\raisebox{5.15pt}{
	\scalebox{.9}{\begin{tikzpicture}
	\draw (0pt,0pt) -- (30pt,0pt);
	\draw (0pt,0pt) -- (20pt,-20pt);
	\draw (0pt,0pt) -- (16pt,-32pt);
	\draw (0pt,0pt) -- (-15pt,-30pt);
	
	\draw (20pt,-20pt) -- (-15pt,-30pt);
	\draw (16pt,-32pt) -- (-15pt,-30pt);
	
	\draw (20pt,-20pt) -- (30pt,0pt);
	
	\draw[dashed, arrows=-stealth] (30pt,0pt) -- (52pt,0pt);
	\draw (20pt,-20pt) -- (40pt,-34pt);
	\draw (16pt,-32pt) -- (36pt,-46pt);
	\draw (0pt,0pt) -- (10pt,20pt);
	
	\draw[dashed] (-5pt,0pt) -- (-34pt,0pt);
	\draw[dashed, arrows=stealth-] (-3pt,0pt) -- (-34pt,7pt);
	\draw[dashed] (-5pt,1pt) -- (-32pt,13.1pt);
	
	\draw[black, fill=white] (-35pt,0pt) circle (0.4ex);
	\draw[fill=black] (-35pt,7pt) circle (0.4ex);
	\draw[fill=black] (-35pt,14pt) circle (0.4ex);
	
	\draw (-38pt,17pt) -- (-40pt,17pt) -- (-40pt,-3pt) -- (-38pt,-3pt);
	\draw (-42pt,7pt) -- (-40pt,7pt);
	
	\draw[black, fill=white] (0pt,0pt) circle (0.6ex);
	\draw[black, fill=white] (30pt,0pt) circle (0.6ex);
	\draw[fill=black] (55pt,0pt) circle (0.6ex);
	\draw[fill=black] (20pt,-20pt) circle (0.6ex);
	\draw[fill=black] (16pt,-32pt) circle (0.6ex);
	\draw[black, fill=white] (-15pt,-30pt) circle (0.8ex);
	\draw[black, fill=white] (-15pt,-30pt) circle (0.6ex);
	
	\fill[black] (38pt,-32pt) rectangle (42pt,-36pt);
	\fill[black] (34pt,-44pt) rectangle (38pt,-48pt);
	\fill[black] (8pt,18pt) rectangle (12pt,22pt);
	\node[anchor=west] at (40pt,-34pt) {\scriptsize $4$};
	\node[anchor=west] at (36pt,-46pt) {\scriptsize $3$};
	\node[anchor=west] at (10pt,21pt) {\scriptsize $x+1$};
	
	\node[anchor=east] at (6pt,8pt) {\scriptsize $A$};
	\node[anchor=west] at (20pt,-19pt) {\scriptsize $B_1$};
	\node[anchor=west] at (16pt,-31pt) {\scriptsize $B_2$};
	\node[anchor=north] at (-12pt,-31pt) {\scriptsize $C$\tiny$^{(3)}$};
	\node[anchor=west] at (28pt,7pt) {\scriptsize $D$};
	\node[anchor=west] at (53pt,6pt) {\scriptsize $v$};
	
	\node[anchor=east] at (-40pt,10pt) {\scriptsize $x$ input};
	\node[anchor=east] at (-42.5pt,3pt) {\scriptsize nodes};
\end{tikzpicture}}
	}
	\caption{\textsc{and} gate}
	\label{fig:and}
\endminipage\hfill
\hspace{0.03\textwidth}
\setcounter{figure}{7}
\minipage{0.26\textwidth}
		\scalebox{.9}{\begin{tikzpicture}
	\draw[dashed, arrows=-stealth] (50pt,0pt) -- (67pt,0pt);
	\draw (50pt,0pt) -- (20pt,41pt);
	\draw (50pt,0pt) -- (20pt,11pt);
	\draw (50pt,0pt) -- (30pt,-10pt);
	\draw (50pt,0pt) -- (20pt,-36pt);
	
	\draw (0pt,41pt) -- (20pt,41pt);
	\draw (0pt,11pt) -- (20pt,11pt);
	\draw (0pt,-36pt) -- (20pt,-36pt);
	
	\draw (50pt,0pt) -- (50pt,20pt);
	\fill[black] (48pt,18pt) rectangle (52pt,22pt);
	
	\draw[dashed, arrows=stealth-] (-4pt,41pt) -- (-25pt,41pt);
	\draw[dashed, arrows=stealth-] (-4pt,11pt) -- (-25pt,11pt);
	\draw[dashed, arrows=stealth-] (-4pt,-36pt) -- (-25pt,-36pt);
	
	\draw (0pt,41pt) -- (-12pt,29pt);
	\draw (0pt,11pt) -- (-12pt,-1pt);
	\draw (0pt,-36pt) -- (-12pt,-48pt);
	
	\draw[fill=black] (70pt,0pt) circle (0.6ex);
	\draw[black, fill=white] (50pt,0pt) circle (0.6ex);
	
	\draw[black, fill=white] (20pt,41pt) circle (0.8ex);
	\draw[fill=black] (20pt,41pt) circle (0.6ex);
	\draw[black, fill=white] (20pt,11pt) circle (0.8ex);
	\draw[black, fill=white] (20pt,11pt) circle (0.6ex);
	\node[anchor=north] at (10pt,-9pt) {\normalsize ...};
	\draw[black, fill=white] (20pt,-36pt) circle (0.8ex);
	\draw[black, fill=white] (20pt,-36pt) circle (0.6ex);
	
	\draw[black, fill=white] (0pt,41pt) circle (0.8ex);
	\draw[black, fill=white] (0pt,41pt) circle (0.6ex);
	\draw[black, fill=white] (0pt,11pt) circle (0.8ex);
	\draw[fill=black] (0pt,11pt) circle (0.6ex);
	\draw[black, fill=white] (0pt,-36pt) circle (0.8ex);
	\draw[fill=black] (0pt,-36pt) circle (0.6ex);
	
	\draw[fill=black] (-25pt,41pt) circle (0.6ex);
	\draw[black, fill=white] (-25pt,11pt) circle (0.6ex);
	\draw[black, fill=white] (-25pt,-36pt) circle (0.6ex);
	\draw[black, fill=white] (-10pt,27pt) rectangle (-14pt,31pt);
	\fill[black] (-10pt,-3pt) rectangle (-14pt,1pt);
	\fill[black] (-10pt,-50pt) rectangle (-14pt,-46pt);
	
	\node[anchor=east] at (-25pt,45pt) {\footnotesize $v_1$};
	\node[anchor=east] at (-25pt,15pt) {\footnotesize $v_2$};
	\node[anchor=east] at (-25pt,-32pt) {\footnotesize $v_p$};
	
	\node[anchor=west] at (-10pt,50pt) {\scriptsize $A_1$\tiny$\!^{(2)}$};
	\node[anchor=west] at (12pt,50pt) {\scriptsize $B_1$\tiny$\!^{(2)}$};
	\node[anchor=west] at (-10pt,20pt) {\scriptsize $A_2$\tiny$\!^{(2)}$};
	\node[anchor=west] at (12pt,20pt) {\scriptsize $B_2$\tiny$\!^{(2)}$};
	\node[anchor=west] at (-7pt,-45pt) {\scriptsize $A_p$\tiny$\!^{(2)}$};
	\node[anchor=west] at (15pt,-45pt) {\scriptsize $B_p$\tiny$\!^{(2)}$};
	\node[anchor=west] at (50pt,5pt) {\scriptsize $C$};
	\node[anchor=west] at (70pt,4pt) {\footnotesize $v$};
	
	\node[anchor=east] at (-11pt,27pt) {\scriptsize 2};
	\node[anchor=east] at (-11pt,-3pt) {\scriptsize 2};
	\node[anchor=east] at (-11pt,-50pt) {\scriptsize 2};
	\node[anchor=west] at (49pt,22pt) {\scriptsize 2};

\end{tikzpicture}}
	\caption{Join gadget}
	\label{fig:join}
\endminipage\hfill
\hspace{0.03\textwidth}
\minipage{0.25\textwidth}
	 \scalebox{.9}{\begin{tikzpicture}
	\draw[dashed, arrows=stealth-] (-3pt,0pt) -- (-20pt,0pt);
	\draw[dashed, arrows=-stealth] (0pt,0pt) -- (28pt,-43pt);
	\draw[dashed, arrows=-stealth] (0pt,0pt) -- (28pt,-28.5pt);
	\draw[dashed] (0pt,0pt) -- (15pt,-7.5pt);
	\draw[dashed] (0pt,0pt) -- (15pt,0pt);
	\draw[dashed, arrows=-stealth] (0pt,0pt) -- (27.5pt,14pt);
	\draw[dashed, arrows=-stealth] (0pt,0pt) -- (28pt,28.5pt);
	\draw[dashed, arrows=-stealth] (0pt,0pt) -- (28pt,43pt);
	
	\draw[dashed, arrows=stealth-] (57pt,-45pt) -- (30pt,-45pt);
	\draw[dashed, arrows=stealth-] (57pt,-30pt) -- (30pt,-30pt);
	\draw[dashed, arrows=stealth-] (57pt,15pt) -- (30pt,15pt);
	\draw[dashed, arrows=stealth-] (57pt,30pt) -- (30pt,30pt);
	\draw[dashed, arrows=stealth-] (57pt,45pt) -- (30pt,45pt);
	
	\draw (50pt,-52pt) -- (30pt,-45pt);
	\draw (50pt,-37pt) -- (30pt,-30pt);
	\draw (50pt,8pt) -- (30pt,15pt);
	\draw (50pt,23pt) -- (30pt,30pt);
	\draw (50pt,38pt) -- (30pt,45pt);
	
	\draw (30pt,-55pt) -- (30pt,-45pt);
	\draw (30pt,-45pt) -- (30pt,-30pt);
	\draw (30pt,-30pt) -- (30pt,-21pt);
	\draw (30pt,15pt) -- (30pt,9pt);
	\draw (30pt,15pt) -- (30pt,30pt);
	\draw (30pt,30pt) -- (30pt,45pt);
	\draw (30pt,45pt) -- (30pt,55pt);
	
	\draw[black, fill=white] (0pt,0pt) circle (0.6ex);
	\draw[fill=black] (30pt,-45pt) circle (0.6ex);
	\draw[black, fill=white] (30pt,-30pt) circle (0.6ex);
	\draw[fill=black] (30pt,15pt) circle (0.6ex);
	\node[anchor=north] at (30pt,-5pt) {\footnotesize ...};
	\draw[black, fill=white] (30pt,30pt) circle (0.6ex);
	\draw[fill=black] (30pt,45pt) circle (0.6ex);
	\draw[black, fill=white] (60pt,-45pt) circle (0.6ex);
	\draw[fill=black] (60pt,-30pt) circle (0.6ex);
	\draw[black, fill=white] (60pt,15pt) circle (0.6ex);
	\draw[fill=black] (60pt,30pt) circle (0.6ex);
	\draw[black, fill=white] (60pt,45pt) circle (0.6ex);
	\draw[black, fill=white] (28pt,-57pt) rectangle (32pt,-53pt);
	\fill[black] (28pt,57pt) rectangle (32pt,53pt);
	
	\fill[black] (48pt,-54pt) rectangle (52pt,-50pt);
	\draw[black, fill=white] (48pt,-39pt) rectangle (52pt,-35pt);
	\fill[black] (48pt,6pt) rectangle (52pt,10pt);
	\draw[black, fill=white] (48pt,21pt) rectangle (52pt,25pt);
	\fill[black] (48pt,36pt) rectangle (52pt,40pt);
	
	\node[anchor=east] at (2pt,6pt) {\scriptsize $v$};
	\node[anchor=west] at (29pt,50pt) {\scriptsize $F_1$};
	\node[anchor=west] at (29pt,35pt) {\scriptsize $F_2$};
	\node[anchor=west] at (29pt,20pt) {\scriptsize $F_3$};
	\node[anchor=west] at (29pt,-25pt) {\scriptsize $F_{q-1}$};
	\node[anchor=west] at (29pt,-40pt) {\scriptsize $F_q$};
	
	\node[anchor=west] at (59pt,50pt) {\scriptsize $v_1$};
	\node[anchor=west] at (59pt,35pt) {\scriptsize $v_2$};
	\node[anchor=west] at (59pt,20pt) {\scriptsize $v_3$};
	\node[anchor=west] at (59pt,-25pt) {\scriptsize $v_{q-1}$};
	\node[anchor=west] at (59pt,-40pt) {\scriptsize $v_q$};
\end{tikzpicture}}
	\vspace{0.1pt}
	\caption{Fork gadget}
	\label{fig:fork}
\endminipage\hfill
\end{figure}

Note that we could achieve this functionality with a single node, by carefully setting its initial balance such that it switches exactly when all inputs have the desired color. However, \textsc{and} gates are used to ``check'' the state of specific nodes in the construction, and as such, it is unfortunate that this check also affects the nodes that are being checked: once the node in this simple \textsc{and} gate switches, the balance of all input nodes in $X$ will increase by 2. It would be much better to have a gadget that can perform this task without having any effect on the nodes in $X$.

For this purpose, consider the gadget in Figure \ref{fig:and}, which is connected to the nodes in $X$ on the input side and a black node $v$ on the output side. Once all nodes in $X$ are white, node $A$ switches, followed by $B_1$ and $B_2$, and then by $C$. With $C$ switched, $A$ decides to switch back to its original color white. However, since now both $A$ and $B_1$ are white, this finally switches $D$ to black, triggering a change in the output node $v$ (Figure \ref{fig:and_phases}). The usefulness of the gadget lies in the fact that once the entire sequence is completed, $A$ is switched back to its original color, and thus the balance of nodes in $X$ is the same at the end of the process as in the beginning.

\setcounter{figure}{6}
\begin{figure}
\centering
	\scalebox{.9}{\input{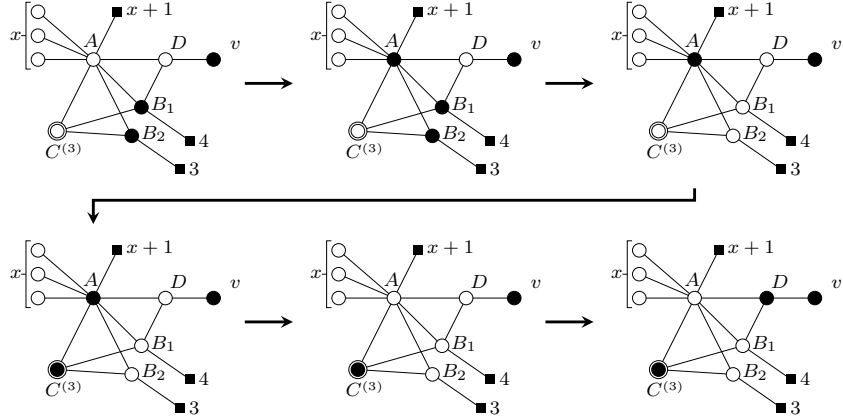}}
	\caption{Operation of an \textsc{and} gate. In the end, node $D$ switches to black, making $v$ switchable.}
	\label{fig:and_phases}
\end{figure}
\setcounter{figure}{9}

\subparagraph*{Join and fork gadgets.}
Finally, we need two small gadgets in the construction to fork and join the control sequence at the ends of our main relay chain.

The \textit{join gadget} of Figure \ref{fig:join} consists of a central node $C$, and an even number of distinct 2-group starter gadgets of alternating color. When an input node $v_i$ switches, then so does $A_i$ and then $B_i$ in the corresponding starter gadget, which also switches $C$ and triggers node $v$. Then when $v_{i+1}$ later switches at some point, the same thing happens to the next starter gadget and $C$ again, only with the two colors swapping roles. Thus if the nodes $v_1$, $v_2$, ... are switched one after another, then each will trigger a new change in the output $v$.

The \textit{fork gadget} of Figure \ref{fig:fork}, on the other hand, is responsible for receiving triggers from a given node $v$, and directing the propagation to a new branch every time.
This is done through a series of nodes $F_i$ of alternating color, each connected to a different output node $v_i$, and also to $F_{i-1}$ and $F_{i+1}$. When $v$ first switches in this setting, only $F_1$ will become switchable, triggering node $v_1$. Similarly, after $v$ is switched for the $i^{\text{th}}$ time, only node $F_i$ becomes switchable, and it triggers the $i^{\text{th}}$ branch of output.

\subparagraph*{Assembling the pieces.}
Our final graph construction (shown in Figure \ref{fig:system}) has two defining parameters $m$ and $r$. The base of the construction is a chain of $m$ rechargeable relays. The leftmost relay's base node is the output node in a join gadget of $r$ branches, while the rightmost relay's base node is an input node in a fork gadget of $r-1$ branches. Finally, for each $i \in \{1, ..., r-1\}$, we add a sequence of gadgets (a \textit{branch}) to connect the $i^{\text{th}}$ output of the fork to the $i+1^{\text{th}}$ input of the join gadget, which is responsible for recharging the relay chain.

Each branch consists of recharging systems connected to our main chain. First let us consider the rechargeable relays where node $U$ is currently white (either the even or the odd ones; relays at positions of the same parity are all in the same state). We first need a recharging system to recharge all these relays, and then we need another system to reset the relays. We need similarly 2 recharging systems for the other half of the relays which are in the opposite color phase.

Finally, we need to force the player to indeed execute these changes on the relays. For that, we insert an \textsc{and} gate after each recharging system, which checks if all switchable nodes have indeed been switched before moving on. The output of the \textsc{and} gate is then used to enable the next recharging systems (or the next input of the join gadget, if this was the last system).

This construction ensures that the player has no other choice than to go through the relay chain, follow the next branch from the fork, recharge and reset all the relays, and start going through the relay chain again. Since the chain consists of $m$ relays and it is traversed $r$ times in this process, the switches in the chain add up to $m \cdot r$ steps altogether.

Of course, one also needs to introduce a starting point (initially switchable node) into the construction. This can be done by replacing $v_1$ in the join gadget by a fixed white node.

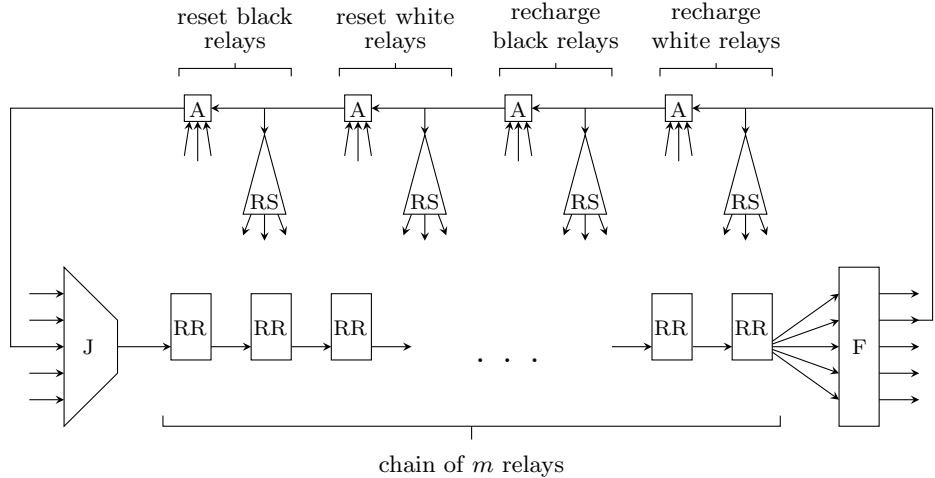
\begin{figure}
\centering
	\begin{tikzpicture}
	
	\draw[arrows=-stealth] (50pt,0pt) -- (50pt,-10pt);
	\draw (50pt,-10pt) -- (58pt,-40pt) -- (42pt,-40pt) -- cycle;
	\draw[arrows=-stealth] (45pt,-40pt) -- (42pt,-48pt);
	\draw[arrows=-stealth] (50pt,-40pt) -- (50pt,-50pt);
	\draw[arrows=-stealth] (55pt,-40pt) -- (58pt,-48pt);
	
	\draw[black, fill=white] (20pt,-5pt) rectangle (30pt,5pt);
	\draw[arrows=-stealth] (20pt,-18pt) -- (22pt,-5pt);
	\draw[arrows=-stealth] (25pt,-20pt) -- (25pt,-5pt);
	\draw[arrows=-stealth] (30pt,-18pt) -- (28pt,-5pt);
	\draw[arrows=-stealth] (80pt,0pt) -- (30pt,0pt);
	
	\node[anchor=south] at (50pt,-42pt) {\scriptsize RS};
	\node[anchor=south] at (25pt,-7pt) {\scriptsize A};
	
	\draw[arrows=-stealth] (110pt,0pt) -- (110pt,-10pt);
	\draw (110pt,-10pt) -- (118pt,-40pt) -- (102pt,-40pt) -- cycle;
	\draw[arrows=-stealth] (105pt,-40pt) -- (102pt,-48pt);
	\draw[arrows=-stealth] (110pt,-40pt) -- (110pt,-50pt);
	\draw[arrows=-stealth] (115pt,-40pt) -- (118pt,-48pt);
	
	\draw[black, fill=white] (80pt,-5pt) rectangle (90pt,5pt);
	\draw[arrows=-stealth] (80pt,-18pt) -- (82pt,-5pt);
	\draw[arrows=-stealth] (85pt,-20pt) -- (85pt,-5pt);
	\draw[arrows=-stealth] (90pt,-18pt) -- (88pt,-5pt);
	\draw[arrows=-stealth] (140pt,0pt) -- (90pt,0pt);
	
	\node[anchor=south] at (110pt,-42pt) {\scriptsize RS};
	\node[anchor=south] at (85pt,-7pt) {\scriptsize A};
	
	\draw[arrows=-stealth] (170pt,0pt) -- (170pt,-10pt);
	\draw (170pt,-10pt) -- (178pt,-40pt) -- (162pt,-40pt) -- cycle;
	\draw[arrows=-stealth] (165pt,-40pt) -- (162pt,-48pt);
	\draw[arrows=-stealth] (170pt,-40pt) -- (170pt,-50pt);
	\draw[arrows=-stealth] (175pt,-40pt) -- (178pt,-48pt);
	
	\draw[black, fill=white] (140pt,-5pt) rectangle (150pt,5pt);
	\draw[arrows=-stealth] (140pt,-18pt) -- (142pt,-5pt);
	\draw[arrows=-stealth] (145pt,-20pt) -- (145pt,-5pt);
	\draw[arrows=-stealth] (150pt,-18pt) -- (148pt,-5pt);
	\draw[arrows=-stealth] (200pt,0pt) -- (150pt,0pt);
	
	\node[anchor=south] at (170pt,-42pt) {\scriptsize RS};
	\node[anchor=south] at (145pt,-7pt) {\scriptsize A};
	
	\draw[arrows=-stealth] (230pt,0pt) -- (230pt,-10pt);
	\draw (230pt,-10pt) -- (238pt,-40pt) -- (222pt,-40pt) -- cycle;
	\draw[arrows=-stealth] (225pt,-40pt) -- (222pt,-48pt);
	\draw[arrows=-stealth] (230pt,-40pt) -- (230pt,-50pt);
	\draw[arrows=-stealth] (235pt,-40pt) -- (238pt,-48pt);
	
	\draw[black, fill=white] (200pt,-5pt) rectangle (210pt,5pt);
	\draw[arrows=-stealth] (200pt,-18pt) -- (202pt,-5pt);
	\draw[arrows=-stealth] (205pt,-20pt) -- (205pt,-5pt);
	\draw[arrows=-stealth] (210pt,-18pt) -- (208pt,-5pt);
	\draw[arrows=-stealth] (293pt,-80pt) -- (300pt,-80pt) -- (300pt,0pt) -- (210pt,0pt);
	
	\node[anchor=south] at (230pt,-42pt) {\scriptsize RS};
	\node[anchor=south] at (205pt,-7pt) {\scriptsize A};
	
	\draw[black, fill=white] (265pt,-60pt) rectangle (280pt,-120pt);
	\draw[arrows=-stealth] (280pt,-70pt) -- (295pt,-70pt);
	\draw[arrows=-stealth] (280pt,-80pt) -- (295pt,-80pt);
	\draw[arrows=-stealth] (280pt,-90pt) -- (295pt,-90pt);
	\draw[arrows=-stealth] (280pt,-100pt) -- (295pt,-100pt);
	\draw[arrows=-stealth] (280pt,-110pt) -- (295pt,-110pt);
	
	\node[anchor=west] at (266pt,-90pt) {\scriptsize F};
	
	\draw[black, fill=white] (225pt,-95pt) rectangle (240pt,-70pt);
	\node[anchor=west] at (222pt,-83pt) {\scriptsize RR};
	
	\draw[arrows=-stealth] (240pt,-88pt) -- (265pt,-70pt);
	\draw[arrows=-stealth] (240pt,-89pt) -- (265pt,-80pt);
	\draw[arrows=-stealth] (240pt,-90pt) -- (265pt,-90pt);
	\draw[arrows=-stealth] (240pt,-91pt) -- (265pt,-100pt);
	\draw[arrows=-stealth] (240pt,-92pt) -- (265pt,-110pt);
	
	\draw[arrows=-stealth] (210pt,-90pt) -- (225pt,-90pt);	
	\draw[black, fill=white] (195pt,-95pt) rectangle (210pt,-70pt);
	\node[anchor=west] at (192pt,-83pt) {\scriptsize RR};
	\draw[arrows=-stealth] (180pt,-90pt) -- (195pt,-90pt);
	
	%now the beginning
	
	\draw[arrows=-stealth] (20pt,0pt) -- (-45pt,0pt) -- (-45pt,-90pt) -- (-25pt,-90pt);
	
	\draw (-25pt,-60pt) -- (-25pt,-120pt) -- (-5pt,-100pt) -- (-5pt,-80pt) -- cycle;
	\node[anchor=west] at (-22pt,-90pt) {\scriptsize J};
	\draw[arrows=-stealth] (-38pt,-70pt) -- (-25pt,-70pt);
	\draw[arrows=-stealth] (-38pt,-80pt) -- (-25pt,-80pt);
	\draw[arrows=-stealth] (-38pt,-100pt) -- (-25pt,-100pt);
	\draw[arrows=-stealth] (-38pt,-110pt) -- (-25pt,-110pt);
	
	\draw[arrows=-stealth] (-5pt,-90pt) -- (15pt,-90pt);
	
	\draw[black, fill=white] (15pt,-95pt) rectangle (30pt,-70pt);
	\node[anchor=west] at (12pt,-83pt) {\scriptsize RR};
	\draw[arrows=-stealth] (30pt,-90pt) -- (45pt,-90pt);
	
	\draw[black, fill=white] (45pt,-95pt) rectangle (60pt,-70pt);
	\node[anchor=west] at (42pt,-83pt) {\scriptsize RR};
	\draw[arrows=-stealth] (60pt,-90pt) -- (75pt,-90pt);
	
	\draw[black, fill=white] (75pt,-95pt) rectangle (90pt,-70pt);
	\node[anchor=west] at (72pt,-83pt) {\scriptsize RR};
	\draw[arrows=-stealth] (90pt,-90pt) -- (105pt,-90pt);
	
	\node[anchor=west] at (125pt,-95pt) {\Large . . . };
	
	\draw (12pt,-116pt) -- (12pt,-120pt) -- (243pt,-120pt) -- (243pt,-116pt);
	\draw (127.5pt,-120pt) -- (127.5pt,-125pt);
	\node[anchor=north] at (127.5pt,-127pt) {\footnotesize chain of $m$ relays };
	
	\draw (18pt,12pt) -- (18pt,15pt) -- (60pt,15pt) -- (60pt,12pt);
	\draw (39pt,15pt) -- (39pt,17pt);
	\node[anchor=south] at (39pt,17pt) {\footnotesize relays};
	\node[anchor=south] at (39pt,28pt) {\footnotesize reset black};
	
	\draw (78pt,12pt) -- (78pt,15pt) -- (120pt,15pt) -- (120pt,12pt);
	\draw (99pt,15pt) -- (99pt,17pt);
	\node[anchor=south] at (99pt,17pt) {\footnotesize relays};
	\node[anchor=south] at (99pt,28pt) {\footnotesize reset white};
	
	\draw (138pt,12pt) -- (138pt,15pt) -- (180pt,15pt) -- (180pt,12pt);
	\draw (159pt,15pt) -- (159pt,17pt);
	\node[anchor=south] at (159pt,17pt) {\footnotesize black relays};
	\node[anchor=south] at (159pt,28pt) {\footnotesize recharge};
	
	\draw (198pt,12pt) -- (198pt,15pt) -- (240pt,15pt) -- (240pt,12pt);
	\draw (219pt,15pt) -- (219pt,17pt);
	\node[anchor=south] at (219pt,17pt) {\footnotesize white relays};
	\node[anchor=south] at (219pt,28pt) {\footnotesize recharge};

\end{tikzpicture}
	\caption{Overview of the construction, with one branch shown in detail. Rechargeable relays (RR), Recharging systems (RS), \textsc{and} gates (A), Joins (J) and Forks (F) are explicitly shown.}
	\label{fig:system}
\end{figure}

Now let us consider the number of nodes in the construction. Since rechargeable relays consist of constantly many nodes, the size of the relay chain is $O(m)$. The size of the join and fork gadgets is $O(r)$. Finally, each of the $r-1$ recharging branches consist of constantly many recharging systems, \textsc{and} gates and simple relays; since the latter two have constant size, branch size is dominated by the size of the recharging systems. Each such system is connected to $\frac{m}{2}$ relays, and thus needs to reduce the balance of $O(m)$ nodes by a constant value of 6. This implies that each recharging system needs $O(\sqrt{m})$ nodes.

This shows that we can choose $r=\Theta(\sqrt{m})$ and $m=\Theta(n)$ for our parameters. Our graph then contains $O(m)+O(r)+r \cdot O(\sqrt{m}) = O(n)$ nodes, so it is indeed a valid setting with the proper choice of constants.

To investigate runtime, it is enough to consider the switches in the main relay chain. Each of the $\Theta(n)$ relays has a base node that is switched $\Theta(\sqrt{n})$ times, adding up to a total of $\Omega(n^{3/2})$ switches.

\begin{theorem}
There exists a graph construction with $\Omega(n^{3/2})$ stabilization time in model B.
\end{theorem}

\vspace{-1pt}

Note that in the previous construction, whenever any of the base nodes of the relay chain are switchable, there is no other switchable node in the entire graph. This implies that even in the independent benevolent case, the player has no other option than to select this single node, so the number of minimal switches is $\Omega(n^{3/2})$ even if we assume the independent benevolent model.

In fact, one can observe that the construction also ensures that regardless of the choices of the player, the set of switchable nodes is always an independent set at any point in the process. This implies that models C and D are in fact the same in this graph, and thus the lower bound also holds for model D. This then implies the same bound for all the remaining models.

\begin{corollary}
There is a graph construction with $\Omega(n^{3/2})$ stabilization time in models C–G.
\end{corollary}

\section{Recursive construction} \label{sec:recurse}

We now briefly outline the modification idea that provides the almost tight lower bound of $\Omega(n^{2-\epsilon})$. A detailed discussion of the construction can be found in Appendix \ref{App:B}.

The key idea is to make the recharging systems themselves also rechargeable. Note that once a recharging system has been used, the color of its nodes is exactly that of a recharging system of the opposite color. Thus, if we reset the balance of each node in the system to its initial value, then we can use the system again to recharge the same output nodes repeatedly.

Therefore, we can add a layer of \textit{second-level} recharging systems to recharge all the original (first-level) systems in the graph after all first-level system have been used (as shown in Figure \ref{fig:recurse} in the appendix). Recall that decreasing the sum of balances in a set of nodes by $\chi$ requires a recharging system of $O(\sqrt{\chi})$ nodes. We have $\Theta(\sqrt{m})$ first-level systems in our graph, each consisting of $\Theta(\sqrt{m})$ nodes with a balance of $\Theta(\sqrt{m})$ after use; thus to reset each node in these systems to their default balance of 1, with $\chi=\Theta(m^{3/2})$, a second-level system requires $\sqrt{\chi}=\Theta(m^{3/4})$ nodes.

In order to keep the overall number of nodes in second-level systems in $O(m)$, we add $\Theta(m^{1/4})$ distinct second-level systems to our graph. When used, each one of these second-level systems recharges all systems on the first level, which in turn allows us to propagate through the main relay chain $\Theta(m^{1/2})$ times again. Thus together, first and second-level systems allow us to traverse the relay chain $\Theta(m^{1/2}) \cdot \Theta(m^{1/4})$ times.

We can continue this technique in a recursive manner. Assume that we have $\Theta(m^{1/(2^i)})$ distinct $i^{\text{th}}$-level systems in the construction, each consisting of $\Theta(m^{1-1/(2^i)})$ nodes (which, therefore, all have a balance of $\Theta(m^{1-1/(2^i)})$ after they have been used). We can then use an $(i+1)^{\text{th}}$-level recharging system to recharge all of these $i^{\text{th}}$-level systems; since we now have $\chi=\Theta(m^{1/(2^i)}) \cdot \Theta(m^{1-1/(2^i)}) \cdot \Theta(m^{1-1/(2^i)}) = \Theta(m^{(2^{i+1}-1)/(2^i)})$, this requires a next level system of $\sqrt{\chi} = \Theta(m^{(2^{i+1}-1)/(2^{i+1})})$ nodes. In order to keep the nodes in this new level also in $O(m)$, we only add $\Theta(m^{1/(2^{i+1})})$ systems to the $(i+1)^{\text{th}}$ level.

Following this recursive pattern, we obtain a construction that allows us to traverse the main relay chain $\Theta(m^{1/2}) \cdot \Theta(m^{1/4}) \cdot \Theta(m^{1/8}) \cdot ... = \Theta(m^{1-\epsilon})$ times altogether. Since the chain consists of $\Theta(m)$ nodes, this leads to a stabilization time of $\Theta(m^{2-\epsilon})$.

With $\Theta(m^{1/(2^i)})$ recharging systems on the $i^{\text{th}}$ level, the recursive setting allows us to add $\Theta(\log \log m)$ levels until the number of systems on a level decreases to constant. As each level contains $O(m)$ nodes altogether, the resulting graph consists of $\Theta(m \log \log m)$ nodes. This allows for a choice of $m=\Theta(\frac{n}{\log \log n})$, yielding a stabilization time of $\Theta(m^{2-\epsilon}) = \Omega(\frac{n^{2-\epsilon}}{(\log \log n)^{2-\epsilon}})$. As we have such a construction for any $\epsilon>0$, the logarithmic factors in this bound can be removed. Similarly to the non-recursive case, this lower bound holds in all of our models, since propagations over the relay chain are still only possible sequentially.

\vspace{3pt}

\begin{theorem}
For any $\epsilon>0$, there exists a graph construction with $\Omega(n^{2-\epsilon})$ stabilization time in models B–G.
\end{theorem}

\newpage

\bibliography{references}

\begin{thebibliography}{10}

\bibitem{aharoniUGP}
Ron Aharoni, Eric~C Milner, and Karel Prikry.
\newblock Unfriendly partitions of a graph.
\newblock {\em Journal of Combinatorial Theory, Series B}, 50(1):1--10, 1990.

\bibitem{SGPclass5}
Cristina Bazgan, Zsolt Tuza, and Daniel Vanderpooten.
\newblock On the existence and determination of satisfactory partitions in a
  graph.
\newblock In {\em International Symposium on Algorithms and Computation}, pages
  444--453. Springer, 2003.

\bibitem{approx0}
Cristina Bazgan, Zsolt Tuza, and Daniel Vanderpooten.
\newblock Complexity and approximation of satisfactory partition problems.
\newblock In {\em International Computing and Combinatorics Conference}, pages
  829--838. Springer, 2005.

\bibitem{SGPclass4}
Cristina Bazgan, Zsolt Tuza, and Daniel Vanderpooten.
\newblock The satisfactory partition problem.
\newblock {\em Discrete applied mathematics}, 154(8):1236--1245, 2006.

\bibitem{SGPsurvey}
Cristina Bazgan, Zsolt Tuza, and Daniel Vanderpooten.
\newblock Satisfactory graph partition, variants, and generalizations.
\newblock {\em European Journal of Operational Research}, 206(2):271--280,
  2010.

\bibitem{applic1}
Olivier Bodini, Thomas Fernique, and Damien Regnault.
\newblock Crystallization by stochastic flips.
\newblock In {\em Journal of Physics: Conference Series}, volume 226, page
  012022. IOP Publishing, 2010.

\bibitem{applic2}
Olivier Bodini, Thomas Fernique, and Damien Regnault.
\newblock Stochastic flips on two-letter words.
\newblock In {\em 2010 Proceedings of the Seventh Workshop on Analytic
  Algorithmics and Combinatorics (ANALCO)}, pages 48--55. SIAM, 2010.

\bibitem{UGPrayless}
Henning Bruhn, Reinhard Diestel, Agelos Georgakopoulos, and Philipp
  Spr{\"u}ssel.
\newblock Every rayless graph has an unfriendly partition.
\newblock {\em Combinatorica}, 30(5):521--532, 2010.

\bibitem{applic4}
Zhigang Cao and Xiaoguang Yang.
\newblock The fashion game: Network extension of matching pennies.
\newblock {\em Theoretical Computer Science}, 540:169--181, 2014.

\bibitem{applic3}
Jacques Demongeot, Julio Aracena, Florence Thuderoz, Thierry-Pascal Baum, and
  Olivier Cohen.
\newblock Genetic regulation networks: circuits, regulons and attractors.
\newblock {\em Comptes Rendus Biologies}, 326(2):171--188, 2003.

\bibitem{Fazli}
MohammadAmin Fazli, Mohammad Ghodsi, Jafar Habibi, Pooya Jalaly, Vahab
  Mirrokni, and Sina Sadeghian.
\newblock On non-progressive spread of influence through social networks.
\newblock {\em Theoretical Computer Science}, 550:36 -- 50, 2014.

\bibitem{PoAframework}
Michal Feldman and Ophir Friedler.
\newblock A unified framework for strong price of anarchy in clustering games.
\newblock In {\em International Colloquium on Automata, Languages, and
  Programming}, pages 601--613. Springer, 2015.

\bibitem{majority}
Silvio Frischknecht, Barbara Keller, and Roger Wattenhofer.
\newblock Convergence in (social) influence networks.
\newblock In {\em International Symposium on Distributed Computing}, pages
  433--446. Springer, 2013.

\bibitem{MajOther2}
Bernd G{\"a}rtner and Ahad~N Zehmakan.
\newblock Color war: Cellular automata with majority-rule.
\newblock In {\em International Conference on Language and Automata Theory and
  Applications}, pages 393--404. Springer, 2017.

\bibitem{MajOther1}
Bernd G{\"a}rtner and Ahad~N Zehmakan.
\newblock Majority model on random regular graphs.
\newblock In {\em Latin American Symposium on Theoretical Informatics}, pages
  572--583. Springer, 2018.

\bibitem{SGPclass1}
Michael~U Gerber and Daniel Kobler.
\newblock Algorithmic approach to the satisfactory graph partitioning problem.
\newblock {\em European Journal of Operational Research}, 125(2):283--291,
  2000.

\bibitem{SGPclass2}
Michael~U Gerber and Daniel Kobler.
\newblock Classes of graphs that can be partitioned to satisfy all their
  vertices.
\newblock {\em Australasian Journal of Combinatorics}, 29:201--214, 2004.

\bibitem{Goles}
Eric Goles and Jorge Olivos.
\newblock Periodic behaviour of generalized threshold functions.
\newblock {\em Discrete Mathematics}, 30(2):187--189, 1980.

\bibitem{hedetniemi}
Sandra~M Hedetniemi, Stephen~T Hedetniemi, KE~Kennedy, and Alice~A Mcrae.
\newblock Self-stabilizing algorithms for unfriendly partitions into two
  disjoint dominating sets.
\newblock {\em Parallel Processing Letters}, 23(01):1350001, 2013.

\bibitem{votingtime}
Dominik Kaaser, Frederik Mallmann-Trenn, and Emanuele Natale.
\newblock On the voting time of the deterministic majority process.
\newblock In {\em 41st International Symposium on Mathematical Foundations of
  Computer Science (MFCS 2016)}, 2016.

\bibitem{majorityW}
Barbara Keller, David Peleg, and Roger Wattenhofer.
\newblock How even tiny influence can have a big impact!
\newblock In {\em International Conference on Fun with Algorithms}, pages
  252--263. Springer, 2014.

\bibitem{KPRanticoor}
Jeremy Kun, Brian Powers, and Lev Reyzin.
\newblock Anti-coordination games and stable graph colorings.
\newblock In {\em International Symposium on Algorithmic Game Theory}, pages
  122--133. Springer, 2013.

\bibitem{switchOnce}
Yuezhou Lv and Thomas Moscibroda.
\newblock Local information in influence networks.
\newblock In {\em International Symposium on Distributed Computing}, pages
  292--308. Springer, 2015.

\bibitem{Ahad2018}
Ahad N~Zehmakan.
\newblock Opinion forming in erd{\"o}s-r{\'e}nyi random graph and expanders.
\newblock In {\em 29th International Symposium on Algorithms and Computation
  (ISAAC 2018)}. Schloss Dagstuhl-Leibniz-Zentrum fuer Informatik, 2018.

\bibitem{stacs}
P{\'a}l~Andr{\'a}s Papp and Roger Wattenhofer.
\newblock Stabilization time in weighted minority processes.
\newblock In {\em 36th International Symposium on Theoretical Aspects of
  Computer Science (STACS 2019)}. Schloss Dagstuhl-Leibniz-Zentrum fuer
  Informatik, 2019.

\bibitem{CA1}
Damien Regnault, Nicolas Schabanel, and {\'E}ric Thierry.
\newblock Progresses in the analysis of stochastic 2d cellular automata: A
  study of asynchronous 2d minority.
\newblock In Lud{\v{e}}k Ku{\v{c}}era and Anton{\'i}n Ku{\v{c}}era, editors,
  {\em Mathematical Foundations of Computer Science 2007}, pages 320--332.
  Springer Berlin Heidelberg, 2007.

\bibitem{CA2}
Damien Regnault, Nicolas Schabanel, and {\'E}ric Thierry.
\newblock On the analysis of “simple” 2d stochastic cellular automata.
\newblock In {\em International Conference on Language and Automata Theory and
  Applications}, pages 452--463. Springer, 2008.

\bibitem{CA3}
Jean-Baptiste Rouquier, Damien Regnault, and {\'E}ric Thierry.
\newblock Stochastic minority on graphs.
\newblock {\em Theoretical Computer Science}, 412(30):3947--3963, 2011.

\bibitem{SGPclass3}
Khurram~H Shafique and Ronald~D Dutton.
\newblock On satisfactory partitioning of graphs.
\newblock {\em Congressus Numerantium}, pages 183--194, 2002.

\bibitem{noUGP}
Saharon Shelah and Eric~C Milner.
\newblock Graphs with no unfriendly partitions.
\newblock {\em A tribute to Paul Erd{\"o}s}, pages 373--384, 1990.

\bibitem{certainityMaj}
Ariel Webster, Bruce Kapron, and Valerie King.
\newblock Stability of certainty and opinion in influence networks.
\newblock In {\em Advances in Social Networks Analysis and Mining (ASONAM),
  2016 IEEE/ACM International Conference on}, pages 1309--1320. IEEE, 2016.

\bibitem{Winkler}
Peter Winkler.
\newblock Puzzled: Delightful graph theory.
\newblock {\em Commun. ACM}, 51(8):104--104, August 2008.

\end{thebibliography}

\newpage

\begin{appendices}

\section{Discussion of the recursive construction} \label{App:B}

While the main idea of the recursive construction has been outlined above, there are numerous details worth discussing for completeness.

Given a used recharging system, we need to restore the balance of $M$ and $U$ to 1 in order to obtain a recharging system of the opposite color. Then by triggering $U$ again, we can use the system to recharge the nodes in $X$ once more. Note that on each level, we have used recharging systems in both color variants. Since a second-level system can only be used to recharge nodes of the same color, every time we recharge all the first-level systems, we in fact need two second-level recharging systems, one of each color.

\begin{figure}
\centering
\captionsetup{justification=centering}
	\begin{tikzpicture}

	\draw (7pt,0pt) -- (25pt,25pt);
	\draw (15pt,0pt) -- (25pt,25pt);
	\draw (23pt,0pt) -- (25pt,25pt);
	\draw (43pt,0pt) -- (25pt,25pt);
	
	\draw[dashed, arrows=-stealth] (7pt,0pt) -- (4pt,-20pt);
	\draw[dashed, arrows=-stealth] (7pt,0pt) -- (9pt,-20pt);
	\draw[dashed, arrows=-stealth] (7pt,0pt) -- (15.5pt,-20pt);
	
	\draw[dashed, arrows=-stealth] (15pt,0pt) -- (11pt,-20pt);
	\draw[dashed, arrows=-stealth] (15pt,0pt) -- (21pt,-20pt);
	
	\draw[fill=black] (4pt,-22pt) circle (0.3ex);
	\draw[fill=black] (10pt,-22pt) circle (0.3ex);
	\draw[fill=black] (16pt,-22pt) circle (0.3ex);
	\draw[fill=black] (22pt,-22pt) circle (0.3ex);
	\draw[fill=black] (28pt,-22pt) circle (0.3ex);
	
	\draw[dashed, arrows=-stealth] (23pt,0pt) -- (23pt,-20pt);
	\draw[dashed, arrows=-stealth] (23pt,0pt) -- (28pt,-20pt);
	\draw[dashed] (23pt,0pt) -- (28pt,-9pt);
	
	\draw[dashed] (43pt,0pt) -- (43pt,-9pt);
	\draw[dashed] (43pt,0pt) -- (41pt,-11pt);
	\draw[dashed] (43pt,0pt) -- (39pt,-9pt);
	
	\draw (25pt,25pt) -- (25pt,50pt);
	
	\draw[dashed, arrows=stealth-] (28pt,50pt) -- (50pt,50pt);
	\draw (0pt,50pt) -- (25pt,50pt);
	
	\draw (0pt,25pt) -- (25pt,25pt);

	\draw[black, fill=white] (7pt,0pt) circle (0.6ex);
	\draw[black, fill=white] (15pt,0pt) circle (0.6ex);
	\draw[black, fill=white] (23pt,0pt) circle (0.6ex);
	\node[anchor=west] at (25.5pt,-2pt) {\small \textit{...}};
	\draw[black, fill=white] (43pt,0pt) circle (0.6ex);
	
	\draw[black, fill=white] (25pt,25pt) circle (0.9ex);
	\draw[fill=black] (25pt,25pt) circle (0.7ex);
	
	\node[anchor=west] at (23pt,32pt) {\scriptsize$\Theta(\!m^{\frac{1}{2}}\!)$};
	
	\draw[black, fill=white] (25pt,50pt) circle (0.6ex);
	
	\draw[fill=black] (50pt,50pt) circle (0.45ex);
	
	\draw[black, fill=white] (-2pt,52pt) rectangle (2pt,48pt);
	
	\node[anchor=east] at (14pt,43.5pt) {\scriptsize$\Theta(\!m^{\frac{1}{2}}\!)$};
	
	\fill[black] (-2pt,27pt) rectangle (2pt,23pt);
	
	\node[anchor=east] at (14pt,18.5pt) {\scriptsize$\Theta(\!m^{\frac{1}{2}}\!)$};

	\draw[dotted] (-15pt,-27pt) -- (55pt,-27pt) -- (55pt,55pt) -- (-15pt,55pt) -- cycle;
	
	\draw[dotted] (70pt,-27pt) -- (95pt,-27pt) -- (95pt,55pt) -- (70pt,55pt) -- cycle;
	
	\draw[dotted] (110pt,-27pt) -- (135pt,-27pt) -- (135pt,55pt) -- (110pt,55pt) -- cycle;
	
	\draw[dotted] (150pt,-27pt) -- (175pt,-27pt) -- (175pt,55pt) -- (150pt,55pt) -- cycle;

	\draw (57pt,90pt) -- (75pt,115pt);
	\draw (65pt,90pt) -- (75pt,115pt);
	\draw (73pt,90pt) -- (75pt,115pt);
	\draw (93pt,90pt) -- (75pt,115pt);
	
	%---
	
	\draw[dashed] (57pt,90pt) -- (55pt,76pt);
	\draw[dashed] (57pt,90pt) -- (60pt,76pt);
	\draw[dashed] (57pt,90pt) -- (65pt,76pt);
	
	\draw[dashed] (65pt,90pt) -- (62pt,76pt);
	\draw[dashed] (65pt,90pt) -- (69pt,76pt);
	
	\draw[dashed] (73pt,90pt) -- (71pt,76pt);
	\draw[dashed] (73pt,90pt) -- (76pt,76pt);
	\draw[dashed] (73pt,90pt) -- (78pt,81pt);
	
	\draw[dashed] (93pt,90pt) -- (94pt,81pt);
	\draw[dashed] (93pt,90pt) -- (91pt,79pt);
	\draw[dashed] (93pt,90pt) -- (89pt,81pt);
	%---
	
	\draw (54pt,78pt) -- (54pt,75pt) -- (96pt,75pt) -- (96pt,78pt);
	\draw[dashed] (75pt,75pt) -- (75pt,60pt);
	
	\draw[dashed, arrows=-stealth] (75pt,60pt) -- (60pt,60pt) -- (60pt,25pt) -- (29pt,25pt);
	\draw[dashed, arrows=-stealth] (75pt,60pt) -- (100pt,60pt) -- (100pt,25pt) -- (83pt,25pt);
	\draw[dashed, arrows=-stealth] (100pt,60pt) -- (140pt,60pt) -- (140pt,25pt) -- (123pt,25pt);
	\draw[dashed, arrows=-stealth] (140pt,60pt) -- (180pt,60pt) -- (180pt,25pt) -- (163pt,25pt);
	
	\draw (75pt,115pt) -- (75pt,140pt);
	
	\draw[dashed, arrows=stealth-] (78pt,140pt) -- (100pt,140pt);
	\draw (50pt,140pt) -- (75pt,140pt);
	
	\draw (50pt,115pt) -- (75pt,115pt);

	\draw[black, fill=white] (57pt,90pt) circle (0.6ex);
	\draw[black, fill=white] (65pt,90pt) circle (0.6ex);
	\draw[black, fill=white] (73pt,90pt) circle (0.6ex);
	\node[anchor=west] at (75.5pt,88pt) {\small \textit{...}};
	\draw[black, fill=white] (93pt,90pt) circle (0.6ex);
	
	\draw[black, fill=white] (75pt,115pt) circle (0.9ex);
	\draw[fill=black] (75pt,115pt) circle (0.7ex);
	
	\node[anchor=west] at (73pt,122pt) {\scriptsize$\Theta(\!m^{\frac{3}{4}}\!)$};
	
	\draw[black, fill=white] (75pt,140pt) circle (0.6ex);
	
	\draw[fill=black] (100pt,140pt) circle (0.45ex);
	
	\draw[black, fill=white] (48pt,142pt) rectangle (52pt,138pt);
	
	\node[anchor=east] at (64pt,133.5pt) {\scriptsize$\Theta(\!m^{\frac{3}{4}}\!)$};
	
	\fill[black] (48pt,117pt) rectangle (52pt,113pt);
	
	\node[anchor=east] at (64pt,108.5pt) {\scriptsize$\Theta(\!m^{\frac{3}{4}}\!)$};
	
	\draw[dotted] (33pt,71pt) -- (106pt,71pt) -- (106pt,145pt) -- (33pt,145pt) -- cycle;

	\node[anchor=north] at (-58pt,30pt) {\small \textit{first-level}};
	\node[anchor=north] at (-58pt,20pt) {\small \textit{systems}};
	
	\node[anchor=north] at (-16pt,120pt) {\small \textit{second-level}};
	\node[anchor=north] at (-16pt,110pt) {\small \textit{system}};

\end{tikzpicture}
	\caption{Connection of a second-level recharging system to first-level recharging systems. For simplicity, only the recharging of group $M$ is shown (node $U$ also has to be recharged).}
	\label{fig:recurse}
\end{figure}
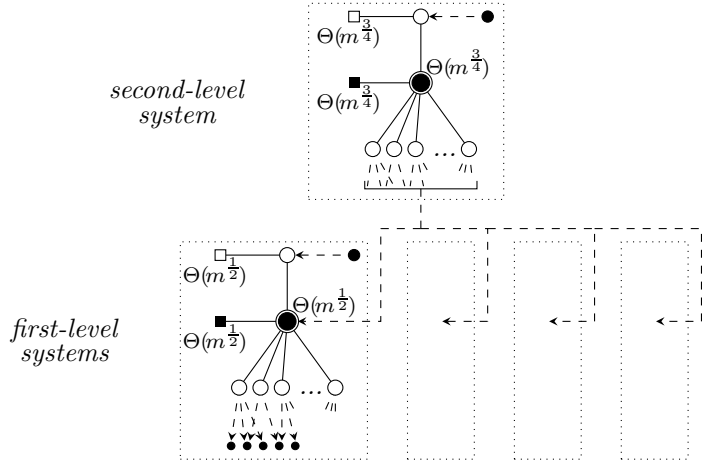

When using one of these second-level systems, it recharges all systems on the first level (of a given color), as shown in Figure \ref{fig:recurse}. This then in turn allows us to propagate through the main relay chain $\Theta(m^{1/2})$ times again. Therefore, with $\Theta(m^{1/4})$ second-level systems in the construction, using only the first two levels, we can already traverse the main chain $\Theta(m^{1/2}) \cdot \Theta(m^{1/4})$ times.

Generally, the recursive construction works the following way. Every time when first-level systems have all been used, an extra branch is added to the construction, which uses one of the second-level systems to recharge the entire first level (and does not influence the relay chain). Similarly, whenever we would need such a second-level branch but all of them has been used, a third-level branch is added to recharge all second-level systems, and the required second-level branch is only visited after traversing this third-level branch.

As described in the summary, continuing this in a recursive manner allows us to propagate through the main relay chain $\Theta(m^{1-\epsilon})$ times for any $\epsilon>0$. If there are $\Theta(m^{1/(2^i)})$ distinct $i^{\text{th}}$-level systems in the graph, and each consist of $\Theta(m^{1-1/(2^i)})$ nodes (which hence all have a balance of $\Theta(m^{1-1/(2^i)})$ after they have been used). An $(i+1)^{\text{th}}$-level recharging system that recharges all these will then have $\chi=\Theta(m^{1/(2^i)}) \cdot \Theta(m^{1-1/(2^i)}) \cdot \Theta(m^{1-1/(2^i)}) = \Theta(m^{(2^{i+1}-1)/(2^i)})$, and therefore needs to consist of $\sqrt{\chi} = \Theta(m^{(2^{i+1}-1)/(2^{i+1})})$ nodes.

As mentioned, we add $\Theta(m^{1/(2^{i+1})})$ such systems in order to keep the nodes on the $(i+1)^{\text{th}}$ level in $O(m)$. The whole recursive construction allows us to go through the main relay chain $\Theta(m^{1/2}) \cdot \Theta(m^{1/4}) \cdot \Theta(m^{1/8}) \cdot ... $ times altogether. If the number of levels go to infinity with $m$ increasing, then for any $\epsilon>0$, there is an $m$ large enough that the number of relay chain traversals is at least $\Theta(m^{1-\epsilon})$. With the length of the relay chain being $\Theta(m)$, this implies a stabilization time of $\Theta(m^{2-\epsilon})$.

With an analysis of the constants in the process, one can show that the coefficient in each factor of $\Theta(m^{1/2}) \cdot \Theta(m^{1/4}) \cdot \Theta(m^{1/8}) \cdot ... $ can be chosen to be at least 1, and thus these constant do not add up to dividing logarithmic factors when taking the product. However, this is in fact unnecessary, as any such logarithmic factor could also be removed simply by a smaller choice of $\epsilon$.

If we have $\Theta(m^{1/(2^i)})$ recharging systems on the $i^{\text{th}}$ level, this setting allows us to add $\Theta(\log \log m)$ levels until the number of systems on a level decreases to a constant value.

Now let us analyze the number of nodes in the graph. On each level, the systems contain $\Theta(m)$ nodes altogether, so the number of nodes in recharging systems adds up to $\Theta(m \log \log m)$ over all levels. The size of the graph is dominated by these nodes. The number of branches controlling first-level systems is $\Theta(m^{1/2} \cdot m^{1/4} \cdot m^{1/8} \cdot ...) = O(m)$, the number of branches controlling second-level systems is only $\Theta(m^{1/4} \cdot m^{1/8} \cdot ...) = O(m^{1/2})$, and so on, the number of $i^{\text{th}}$-level branches is $O(m^{1/2^{i-1}})$. Summing these up, the number of branches altogether is still $O(m)$. Apart from recharging systems, each branch contains constantly many nodes only (in the form of simple relays, \textsc{and} gates, and the corresponding parts of the fork and join gadgets). This shows that the number of nodes outside of the recharging system is only $O(m)$ altogether, thus the number of nodes in the entire graph is indeed $\Theta(m \log \log m)$.

This validates our choice of $m=\Theta(\frac{n}{\log \log n})$, confirming that the stabilization time of the recursive construction is indeed $\Omega(\frac{n^{2-\epsilon}}{(\log \log n)^{2-\epsilon}})$. Since this bound holds for any $\epsilon>0$, we can easily remove the logarithmic factors: a lower bound of $\Omega(n^{2-\epsilon})$ follows from the same construction for any $\widehat{\epsilon} <\epsilon$.

It only remains to discuss some details of the construction. Recall that in addition to the group $M$, the balance of node $U$ also has to be reset between two uses of a recharging system; however, we did not point this out when calculating the necessary size of systems, since besides $M$, a single extra node does not affect the magnitude. Earlier, we have noted that the recharging systems on a certain level consist of two classes of systems of different color; observe that the next level systems that recharge the groups $M$ in one class can simultaneously be used to also recharge the nodes $U$ in the other class. Alternatively (for simpler analysis), we can add an extra recharging system (of the same size) on each branch in order to separately recharge the nodes $U$ on the level below.

Finally, note that in this recursive setting, recharging systems are slightly modified in the sense that they have multiple input nodes from multiple different branches, each connected to node $U$. However, this does not modify the behavior of $U$ as long as its initial balance is readjusted to 1. This also requires a minor modification in the simple relays that are used as input nodes, since relays generally assume that their output node never switches before the relays themselves are triggered. This can be resolved by using a modified relay where the base node has an initial balance of 3, and thus it is enabled by two distinct simple relays on the branch.

\section{Detailed analysis of gadgets} \label{App:A}

Here we provide a more detailed description of the gadgets, and also comment on their behavior and their use in the construction.

\subparagraph*{Simple relay.} The construction and behavior of the simple relay has already been described above. One thing to note is that in our construction, simple relays are always used only once: after node $B$ switches, propagation never returns to the same part of the graph again, and thus node $B$ will remain unswitchable for the rest of the process.

While we mostly use this original version of the gadget, we occasionally need relays that have multiple output nodes instead of just one. Note that this only requires a simple modification: besides connecting $x$ extra (black) output nodes to node $B$, we also need to add $x$ fixed (white) nodes in order to keep the initial balance of $B$ unchanged.

As mentioned, chains of simple relays are mostly used to connect more complex gadgets in our construction. Note that depending on whether the input and output nodes in these gadgets are supposed to have the same or different initial colors, we only need a chain of length 1 or 2 for this, respectively.

\subparagraph*{Rechargeable relay.} In a rechargeable relay, node $B$ is connected to an upper node $U$ instead of a fixed node. Node $U$ is connected to a group $C$ of two nodes, which is further connected to nodes $R_1$, $R_2$. Initially, $C$ has the opposite color as $B$ and $U$, and one of $R_1$ and $R_2$ is white, the other is black. Node $B$ has the same external neighbors as a simple relay. The recharge nodes can both have any set of external neighbors as long as their initial balance is 3 with $C$ ignored. With $C$ also considered, the initial balance of $R_1$ and $R_2$ is then 1 and 5, respectively.

Note that since $R_1$ and $R_2$ have opposite colors, this recharging process can always be executed on a used relay through either $R_1$ or $R_2$, depending on the current color of the nodes. We only need to select the recharge node that has the current color of $U$, and switch 3 of its neighbors (to $U$'s current color) for the recharging step, and then switch 3 of its neighbors (to the opposite color) for the resetting step.

\subparagraph*{Recharging system.} In a basic recharging system, the node $U$ is connected to the input node $v$, the group $M$, and to $\sqrt{n}+1$ fixed white nodes. The middle level group $M$ has a further edge to all nodes $L_i$, and is balanced by $\sqrt{m}$ fixed black nodes. Finally, each node $L_i$ has $\sqrt{m}$ distinct neighbors in $X$, and thus each node in $X$ is connected to exactly one lower-level node. For convenience, we assume that $m$ is a square number.

A generalized recharging system is almost identical to this, except for the nodes $L_i$ occasionally being connected to the same node. The connections between the lower level and $X$ are not directly specified: we are free to choose which of the nodes $L_i$ to connect to a specific node in $X$. Note, however, that the gadget design implicitly assumes that $x_j \leq \sqrt{\chi}$ for all nodes in $X$. This is naturally satisfied whenever we use the gadget in our constructions, since we always have $x_1=x_2=...=x_m$ with $|X| > x_j$. Also note that for convenience, we assume $\chi$ to be a square number.

Nodes in the upper and lower levels are initially white, while $M$ and the input node $v$ are initially black. The nodes $X$ may assume any color, and also may switch multiple times before the recharging system is activated. However, the graph construction ensures that at the time when the gadget is activated (that is, when $v$ switches), all nodes in $X$ are currently colored black (i.e., we indeed use the system on rechargeable relays that can currently be recharged). The gadget design ensures that $U$ and $M$ have an initial balance of 1, while the nodes $L_i$ have a balance of 1 at least, depending on the current color of their neighbors in $X$.

\subparagraph*{AND gate.} The \textsc{and} gate consist of 7 nodes, three of which form a group ($C$). The input nodes of $X$ are connected to node $A$, which is further connected to all other nodes in the gadget ($B_1$, $B_2$, $D$ and the group $C$). Nodes $B_1$ and $B_2$ are also connected to group $C$, node $B_1$ has an edge to node $D$, and node $D$ is connected to some external black node $v$ on the output side. Furthermore, $A$, $B_1$ and $B_2$ have $x+1$, $4$ and $3$ fixed black neighbors, respectively.

One can check that each node has a positive balance as long as there exists a black node in $X$. Node $A$ gets a balance of $x-1$ from the nodes within the gadget, so it is not switchable unless all nodes in $X$ are white. Nodes $B_1$, $B_2$, $C$ and $D$ all have an initial balance of 1.

After the gadget reaches its final stage (see Figure \ref{fig:and_phases}), no node in the gadget can ever change again, regardless of the states of $X$ or $v$.

Note that for the described behavior of the gadget, we also need the fact that none of the nodes in $X$ switch between the first and second switching of $A$. The switching of $A$ only increases their balance (temporarily), so this is guaranteed if other neighbors of nodes in $X$ do not interfere with the process. In the construction, we only use \textsc{and} gates this way: whenever a node $A$ becomes switchable in a gate, then that is the only switchable node in the entire graph, so no other nodes will switch until the propagation reaches $v$.

As long as this condition is fulfilled, we can connect any number of \textsc{and} gates on a given node of the graph without affecting its behavior; we only have to make sure that we also add fixed node neighbors to restore the node's balance to the original value.

\subparagraph*{Join gadget.} A join gadget consists of $p$ distinct starter gadgets, where $p$ is assumed to be an even number. Each starter gadget consists of two groups $A_i$ and $B_i$, both of size 2 (with $i \in \{1,2, ..., p\}$). The two groups are connected to each other, and $A_i$ has a further edge to the input node $v_i$, and two fixed nodes of the same color as its own. Finally, all $B_i$ are connected to a central node $C$, which is in turn connected to an output node $v$. Node $C$ also has two further fixed black connections.

Initially, $A_i$ for odd $i$ values, $B_i$ for even $i$ values, $v_i$ for even $i$ and node $C$ are colored white; the remaining nodes are colored black. Nodes $A_i$ have an initial balance of 1, nodes $B_i$ have an initial balance of 1 or 3 (depending on parity), and $C$ has an initial balance of 3.

As described, the switching of $v_i$ triggers a switch in $A_i$, then $B_i$, then $C$ and finally $v$. After $v$ switches, the balance of $C$ returns to its initial value of 3, so the switching of the next input node will trigger the same process through the next starter gadget.

\subparagraph*{Fork gadget.} The fork gadget consist of $q$ nodes $F_1$, ..., $F_q$, where we assume $q$ to be an odd number. All $F_i$ are connected to the same input node $v$, and each to a distinct output node $v_i$. They are also linked to each other, with $F_i$ connected to $F_{i+1}$ for all $i \in \{1,2, ..., q-1\}$. Also, node $F_1$ and $F_q$ have a fixed neighbor colored black and white, respectively (imitating the role of the nonexistent nodes $F_0$ and $F_{q+1}$). Finally, each $F_i$ has a further fixed neighbor of its original color. Initially, $F_i$ is colored black for odd $i$ and white for even $i$ values.

The balance of $F_1$ and all white $F_i$-s is originally 1 in this setting, while the balance of black $F_i$-s (except for $F_1$) is 3. Hence when $v$ first switches, only $F_1$ will become switchable (and switching it will propagate on through $v_1$). The next time $v$ switches, it switches back to white; with $v$ and $F_1$ both white, $F_2$ can now switch too. The pattern continues all the way to $F_q$: as $F_{i-1}$ has already been switched before, as soon as $v$ switches back to the color of $F_i$, $F_i$ becomes switchable, too, enabling propagation on the next branch. After $v_i$ switches (and remains that way), $F_i$ is not switchable anymore, since $v_i$, $F_{i-1}$ and its fixed neighbor all have the opposite color.

Note that since each switching $F_i$ increases the current balance of $v$ from 1 to 3, we need to switch two neighbors of $v$ in each turn to make $v$ switchable again. This is exactly what happens when $v$ is the base node of the rightmost relay in the chain: between every consecutive switches of $v$, we switch both node $U$ (by the recharging step) and node $v_L$ (by propagation through the chain) in the relay, and thus $v$ becomes switchable again.

Note that since it is connected to the fork gadget, the rightmost rechargeable relay in the chain is a modified one in the sense that its base node has not one, but $q$ right-side neighbors, colored in alternating fashion. However, this fact does not change its behavior at all. The initial balance of the base node is still 1, and every time after $v$ switches, it has one of its neighbors $F_i$ switching in the opposite direction. That has exactly the same effect as if the right neighbor was simply a subsequent relay in the chain, triggered by $v$.

\subparagraph*{On the whole construction.} For convenience, we assume in the construction that both $m$ and $r$ are even numbers.

Recharging systems and \textsc{and} gates, as all other gadgets, are available in two color variants; in the overview of the construction, we did not discuss which variant is used in which case. However, the current state of each relay in each round is straightforward to calculate, so the necessary color of all recharging systems and \textsc{and} gates can easily be determined.

Also, we have seen that \textsc{and} gates are used to ensure that the given recharging or resetting operations have completely been executed. In order to achieve this, in case of the first systems (which recharge relays), the input edges of the gates can be connected to the upper nodes of the corresponding relays, since that is the last node to switch in the sequence. In case of the systems that reset relays, the aim is only to switch the corresponding recharge node of the relay, so we can connect the gates to the recharge nodes.

However, as each \textsc{and} gate belongs to a certain branch of the construction, we also have to ensure that the \textsc{and} gate is only activated when the propagation reaches this branch, and stays inactive as long as previous branches are being processed. Therefore, besides the specified nodes in the relays, the final input node of the \textsc{and} gate is the node which was used to enable the recharging system in question (node $v$ of Figure \ref{fig:recharger}). This way, the gates ensure that \textit{after} the recharging system is activated, propagation only continues if all the resulting switches were executed.

\subsubsection*{Generalization to $\omega(1)$ colors}

One can observe that in the construction of Section \ref{sec:seqben}, except for nodes $A$ in the \textsc{and} gates, all nodes in the graph have a degree of $O(\sqrt{n})$. We can slightly modify the construction and replace each of these \textsc{and} gates with two levels of such gates, with $\Theta(\sqrt{n})$ distinct gates on the first level (each with $\Theta(\sqrt{n})$ input nodes), and a final gate that connects the outputs of these first-level gates. This gives us a construction with the same properties, but a maximum degree of $O(\sqrt{n})$.

This allows us to generalize the lower bound of $\Omega(n^\frac{3}{2})$ to the case of not only $O(1)$, but up to $O(\sqrt{n})$ colors. The technique for this is the same as in the case of $O(1)$ colors: we add a multipartite graph colored with the additional colors, and connect each of its nodes to each original node. With $\Delta=O(\sqrt{n})$ established, it suffices to have $\Theta(\sqrt{n})$ nodes in each of the color classes. Therefore, using only $\Theta(n)$ additional nodes, we can extend the graph by a multipartite graph on $\Theta(\sqrt{n})$ color classes, each consisting of only $\Theta(\sqrt{n})$ nodes.

\section{Notes on simulations} \label{App:C}

Due to its complexity, we have also verified the correctness of the non-recursive construction of Section \ref{sec:seqben} through implementing it and running a simulation of the minority process. Note that in general, it is difficult to simulate a minority process in a benevolent model, since all possible switching sequences would have to be examined to find the one with the smallest number of steps.

Fortunately, the task is significantly simpler in our case, due to the properties of the construction. The key observation in our graph is that whenever propagation is split into multiple parallel threads (that is, when there are multiple switchable nodes at the same time), then propagation on any of these threads does not influence propagation on other threads at all. Specifically, the nodes on separate threads do not have common neighbors except for the beginning and end of such threads; i.e. when a switching node splits the propagation to multiple threads, or when threads are joined in an \textit{\textsc{and}-like} fashion, meaning that a common neighbor only becomes switchable when propagation has been finished in all of the threads. This implies that throughout the process, these threads can be handled completely independently from each other, and the order in which they are processed is irrelevant. Note that this is also the property of the construction which ensures that the set of switchable nodes is an independent set in any state.

If we exploit this property, the process can be simulated easily by always choosing an arbitrary one of the switchable nodes in the graph, knowing that the choice of nodes will not influence the outcome. To verify correctness in such a simulation, we only have to check that in each step of the process, the set of nodes that become switchable is exactly the set of nodes determined by the analysis. Note that the opposite does not happen in our construction: the switching of a node never makes another switchable node unswitchable (this would also contradict the property that switchable nodes form an independent step in any state).

When examining concrete instances of our construction, we used the parameter $r$ as the input to determine the size of the instance. For a given input value of $r$ (always an even number), we have chosen $m=2 \cdot (r-1)^2$, which fits our preconditions on both magnitudes and parity. All other details of the construction are already determined above; the only additional thing to note is that whenever different gadgets are connected through a chain of simple relays, we always use the smallest possible such chain in the implementation.

The simulations verified that the analysis of the construction is correct, and thus stabilization time is indeed $\Omega(n^{3/2})$ in model B. Table \ref{tab:sim} illustrates the number of steps for some choices $r$, along with the resulting number of nodes in the construction. One can observe that the number of steps indeed grows superlinearly in $n$.

\vspace{15pt}

\begin{table}[!htbp]
\centering
\begin{tabular}{ r r r r r }
\hline
\toprule
\multicolumn{1}{c}{Input ($r$)} & & \multicolumn{1}{c}{Nodes ($n$)} & & \multicolumn{1}{c}{Steps} \\[1.5pt]
\hline
 \rule{0pt}{2.55ex} 2 & \hspace{3pt}  & 99 & \hspace{12pt} & 112 \\
 4 &   & 469 &   & 772 \\
 8 &   & 1 929 &   & 5 884 \\
 16 &  & 7 729 &   & 47 404 \\
 24 &   & 17 369 &   & 161 372 \\
 30 &   & 27 119 &   & 316 568 \\
 40 &   & 48 169 &   & 754 108 \\
 60 &   & 108 269 &   & 2 559 188 \\
 80 &   & 192 369 &   & 6 084 268 \\
 100 &   & 300 469 &   & 11 905 348 \\
 120 &   & 432 569 &   & 20 598 428 \\
\bottomrule
\end{tabular}
\captionsetup{justification=centering}
\caption{Number of steps on some specific graphs}
\label{tab:sim}
\end{table}

\end{appendices}

\end{document}